# The time course of visuo-semantic representations in the human brain is captured by combining vision and language models

Boyan Rong [1,2,3,*], Alessandro Thomas Gifford [1], Emrah Düzel [2,3], Radoslaw Martin Cichy [1]

[1] Department of Education and Psychology, Freie Universität Berlin, Berlin, Germany

[2] Institute of Cognitive Neurology and Dementia Research, Otto-von-Guericke-Universität Magdeburg, Magdeburg, Germany

[3] German Center for Neurodegenerative Diseases (DZNE), Otto-von-Guericke-Universität Magdeburg, Magdeburg, Germany

Corresponding Author: Boyan Rong (boyanr.nj@gmail.com)

**The human visual system provides us with a rich and meaningful percept of the world, transforming retinal signals into visuo-semantic representations. For a model of these representations, here we leveraged a combination of two currently dominating approaches: vision deep neural networks (DNNs) and large language models (LLMs). Using large-scale human electroencephalography (EEG) data recorded during object image viewing, we built encoding models to predict EEG responses using representations from a vision DNN, an LLM, and their fusion. We show that the fusion encoding model outperforms encoding models based on either the vision DNN or the LLM alone, as well as previous modelling approaches, in predicting neural responses to visual stimulation. The vision DNN and the LLM complemented each other in explaining stimulus-related signal in the EEG responses. The vision DNN uniquely captured earlier and broadband EEG signals, whereas the LLM uniquely captured later and low frequency signals, as well as detailed visuo-semantic stimulus information. Together, this provides a more accurate model of the time course of visuo-semantic processing in the human brain.**

# 1. Introduction

Visual perception provides us with a rich and detailed understanding of the world around us. This fundamental cognitive process is mediated by the rapid transformation of neural representations across the visual processing hierarchy from simple visual features such as edges into complex visuo-semantic formats carrying meaning [1–8]. Providing a model predictive of these representations is a key step towards understanding vision. Over the last decade, deep neural networks (DNNs) trained on vision tasks such as object categorization have emerged as powerful models to predict visual brain activity [9–12]. However, while the representations of these vision DNNs best predict early and intermediate visual processing stages in the brain [13–19], later stages are less well captured, thus necessitating models with a different type of representational format.

In response to this shortcoming, recent progress came from computational models of human language that capture visuo-semantic information [20–23]. Both vision DNNs trained with language co-supervision (called multimodal DNNs [24,25]), as well as large language models (LLMs) [26–28], surpass the predictive power of vision DNNs in accounting for brain responses to visual stimulation, particularly for higher visual cortical areas [21,29–31]. However, these studies focused on functional magnetic resonance imaging (fMRI) data that lacks the millisecond precision at which visual processing occurs. Thus, they were agnostic to the dynamics of visual processing that are uniquely explained by either vision DNNs or LLMs.

Based on these recent developments, here we hypothesized that combining a vision DNN with a LLM would yield better neural predictions than either model component alone, and that the two components would uniquely predict earlier and later stages of visual processing, respectively.

To test these hypotheses, we predicted temporally resolved human electroencephalography (EEG) responses for thousands of naturalistic images of objects [32] using encoding models [9,33,34] that fused the representations of a vision DNN and an LLM. Confirming our first hypothesis, we found that the fusion encoding model outperformed encoding models trained on the vision DNN or LLM in isolation in predicting EEG responses to visual stimulation. Confirming our second hypothesis, we found that the vision DNN component of the fusion encoding model uniquely captured earlier and broadband EEG signals, whereas the LLM component uniquely captured later and low frequency signals, as well as detailed visuo-semantic stimulus information.

Together, our results provide a modelling approach accounting for the time course of visuo-semantic processing in the human brain.

# 2. Results

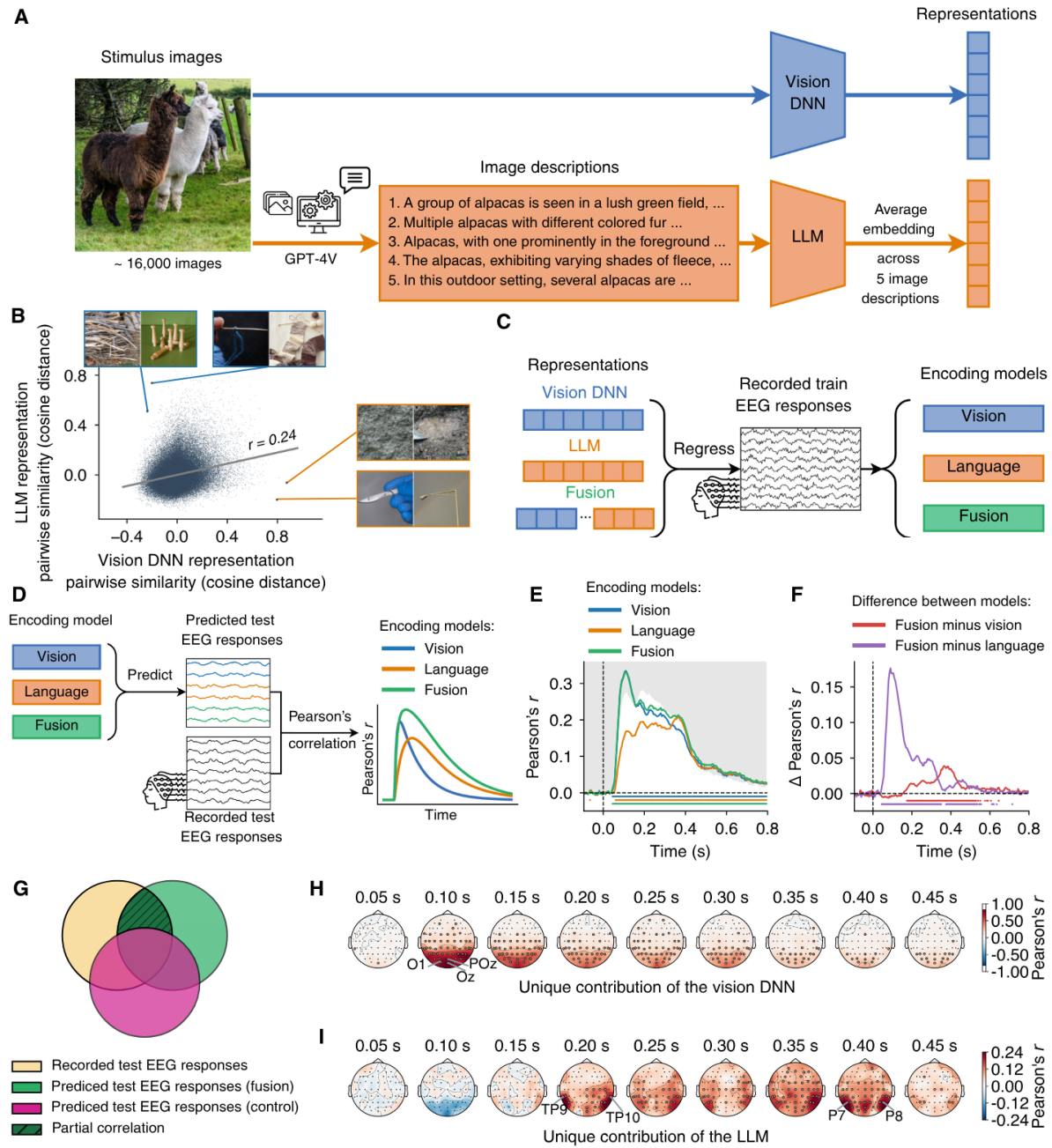

**Figure 1: Combining a vision DNN with an LLM improves the prediction of neural responses to visual stimulation**. **A,** Extraction of vision DNN and LLM representations for each of the 16,740 stimulus images. To obtain the vision DNN representations, we fed each image to the vision DNN and extracted its activations. To obtain the LLM representations, for each image we generated five text descriptions using GPT-4V, independently fed these descriptions to the LLM, and averaged the resulting five embedding instances. **B,** Scatterplot of pairwise cosine similarities between pairs of stimulus representations of either the vision DNN or the LLM. The gray line indicates the linear fit between the vision DNN and LLM pairwise similarities. Inset images are illustrative examples of image pair outliers. **C,** Encoding models training pipeline. We trained encoding models to predict empirically-recorded EEG responses based on representations from vision DNNs, LLM, and their combination. This resulted in three types of encoding models: vision, language, and fusion. **D,** Encoding models testing pipeline. We used the trained encoding models to predict EEG responses to the test stimulus images, and compared (Pearson's *r*) these predictions to the corresponding empirically-recorded EEG responses, resulting in prediction

accuracy time courses. **E,** Prediction accuracy (Pearson's *r*) timecourse for the vision, language, and fusion encoding models. The prediction accuracies are averaged across all participants and EEG channels. In gray is the area between the noise ceiling lower and upper bounds. **F,** Difference in prediction accuracy between the fusion and the vision or language encoding model. **E-F,** The black dashed vertical lines indicate the onset of stimulus presentation, and the black dashed horizontal lines indicate the chance level of no experimental effect. Rows of asterisks at the bottom of the plots indicate significant time points (one-sided *t*-test, $p < 0.05$, FDR corrected across 180 time points, $N = 10$ participants). **G,** Partial correlations between the recorded EEG test responses and the predicted EEG test responses from the fusion encoding model, controlling for the variance explained by the predicted EEG test responses from either the language encoding model (thus isolating the unique contribution of the vision DNN), or from the vision encoding model (thus isolating the unique contribution of the LLM). **H,** EEG topography of partial correlation results, indicating the unique prediction accuracy contribution of the vision DNN. **I,** EEG topography of partial correlation results, indicating the unique prediction accuracy contribution of the LLM. **H-I,** The highlighted black dots indicate significant channels (one-sided *t*-test, $p < 0.05$ FDR-corrected across 63 channels and 180 time points, $N = 10$ participants).

Our core hypotheses were that human neural responses to visual stimulation are better modelled by combining representations from a vision DNN and an LLM than by the representations from either of the two components alone, and that the vision DNN and LLM component would uniquely predict earlier and later stages of visual processing, respectively. The rationale is that vision DNNs and LLMs take complementary roles in predicting neural responses. Vision DNNs capture visual characteristics of visual stimuli, such as color and object shape, and are thus good predictors of earlier stages of neural processing [16,35–37]. Instead, LLMs capture semantic characteristics of visual stimuli such as conceptual and functional meaning, and are thus good predictors of later stages of neural processing [21,38,39].

To test these hypotheses, we combined the vision DNN and the LLM into a fusion encoding model [9,33,34] that predicted neural responses from a large-scale dataset of human EEG recordings to 16,740 naturalistic stimulus images [32]. The vision DNN consisted in CORnet-S [40], a recurrent convolutional DNN trained on object categorization; the LLM consisted in OpenAI's text-embedding-3-large, an LLM trained on vast text corpora.

As a precondition for the fusion encoding model to better predict neural responses than encoding models based on the vision DNN or the LLM components alone, these two components should differently represent information about the same stimulus images. To ascertain this, we extracted stimulus-wise representations by feeding the stimulus images to the vision DNN, and the stimulus image text description (generated by GPT-4V; number of words per image: mean = 26.22, SD = 6.92) to the LLM (**Fig. 1A**). Next, taking 1,000,000 random draws of stimulus pairs from the full stimulus set, we computed the pairwise cosine similarity of the corresponding vision DNN or LLM representations (**Fig. 1B**) [29]. The stimulus pairwise cosine similarities of the vision DNN and LLM representations were moderately correlated (Pearson's $r = 0.24$). This partial representational overlap is expected, as visually similar objects tend to be semantically similar, too [41–43]. However, the moderate correlation indicates that most of the representational content differed between vision DNN and the LLM. To illustrate this, we consider outliers in the correlation plot (i.e., the upper left and lower right quadrant), identifying cases of image pairs that were similarly represented in only the visual DNN or LLM, indicating differences in representational content between the two. Image pairs with low similarity in the vision DNN but high similarity in the LLM (upper left quadrant, e.g., twig and wooden peg, needle and knitting), were visually dissimilar, but conceptually similar. In contrast, image pairs with high similarity in the vision DNN but low similarity in the LLM (lower right quadrant, e.g., streetlight and scalpel, granite and mud) were visually similar, but conceptually dissimilar. Together this shows that the vision DNN and the LLM differ in their representational content, warranting an investigation of their combined representations to predict neural responses to visual stimuli.

## 2.1. Combining a vision DNN with an LLM improves the prediction of neural responses to visual stimuli

To predict human neural responses to visual stimuli from model representations, we trained encoding models using time-resolved EEG responses to over 16,000 images recorded in 10 human participants [32] (**Fig. 1C**). We build three types of encoding models that differed in the format of the stimulus representations used to predict the EEG responses: i) the vision DNN representations (resulting in the vision encoding model, henceforth color-coded in blue); ii) the LLM representations (resulting in the language encoding model, in orange); and iii) the combination of the vision DNN and LLM representations (resulting in the fusion encoding model, in green). We then evaluated the performance of each encoding model type using a set of 200 test images not used for model training, resulting in time courses of prediction accuracy (**Fig. 1D**).

Here and throughout the manuscript, for the participant-averaged time courses of prediction accuracy we assessed significance using *t*-tests (one sided *t*-test, $p < 0.05$, FDR-corrected for 180 time points, $N = 10$ participants). We reported peak latencies with 95% confidence intervals in parenthesis (determined by bootstrapping the participant samples through 10,000 iterations).

All three encoding model types significantly predicted EEG responses to visual stimuli, albeit with different dynamics (**Fig. 1E**). The vision encoding model's prediction accuracy (blue curve) peaked at 110 ms (105 – 115 ms) while the language encoding model's accuracy (orange curve) peaked at 365 ms (185 – 370 ms), with a significant peak latency difference of 255 ms (75 – 265 ms, $p < 10^{-3}$). This suggests that the vision and language encoding models predict earlier and later stages of visual processing, respectively. At their respective peaks, the encoding models cross the lower bounds of the noise ceilings (gray area) indicating that, at these time points, they predict all the explainable signal in the EEG responses.

The fusion encoding model (green curve) outperformed both vision and language encoding models. To ascertain this, we subtracted their prediction accuracies (**Fig. 1F**). The difference in prediction accuracy between the fusion and the language encoding models (purple curve), which isolates the effect of the vision DNN, peaked at 90 ms (90 – 105 ms). In contrast, the difference in prediction accuracy between the fusion and the vision encoding models (red curve), which isolates the effect of the LLM, peaked at 365 ms (360 – 400 ms), with a significant peak latency difference of 275 ms (265 – 310 ms, $p < 10^{-4}$). This confirms our hypotheses that the fusion encoding model outperforms both the vision and the language encoding models, and that the vision and language encoding models predict earlier and later stages of visual processing, respectively.

We established the robustness of these findings through several complementary analyses. First, the result pattern replicated in every single participant (**Suppl. Fig. 1**). Second, we obtained similar results when assessing the encoding models' prediction accuracies with pairwise decoding **(Suppl. Fig. 2)**. Third, we demonstrated the independent contribution of vision and language encoding models in predicting EEG responses also using partial correlation (**Suppl. Fig. 3**) and variance partitioning (**Suppl. Fig. 4**) (see **Suppl. Table 1** for a detailed comparison of onset and peak latency between models).

To estimate the cortical sources uniquely predicted by the vision DNN or the LLM we partialled out the contribution of the language or vision encoding models on the fusion encoding model, respectively, and inspected the resulting partial correlation topographies in EEG sensor space (**Fig. 1G**). The unique contribution of the vision DNN consisted in early prediction accuracy peak at 90 ms (85 – 90 ms) for medial occipito-parietal electrodes (i.e., Oz, POz, O1) (**Fig. 1H**), whereas the unique contribution of the LLM consisted in later prediction accuracies that extended more anteriorly, with bilateral peaks in bilateral occipito-temporal electrodes (i.e., TP9, TP10, P9, P10) (**Fig. 1I**). A closer inspection of the topography of the unique contribution of the language encoding model (**Fig. 1I**) revealed a finer picture, suggesting two distinct processing stages: an early stage with a peak at 200 ms (195 – 260 ms) limited to bilateral temporo-occipital electrodes, and a second stage with a second peak at 365 ms (360 – 400 ms) additionally involving parieto-frontal electrodes. Both peaks occurred later than the peak of the vision encoding model, with a peak difference of 110 ms (105 – 175 ms, $p < 10^{-3}$) and 275 ms (275 – 310 ms, $p < 10^{-4}$), respectively. This suggests two cortically distinct stages of processing captured by the language encoding model. We observed an equivalent topological pattern when using subtraction (**Suppl. Fig. 5**) or variance partitioning (**Suppl. Fig. 6**), further demonstrating the robustness of these findings.

Together, these results confirm our core hypotheses that combining a vision DNN with an LLM improves the prediction of neural responses to visual stimuli compared to using each component in isolation, and that the vision DNN or LLM components uniquely capture earlier versus later stages of EEG responses to visual stimuli, respectively.

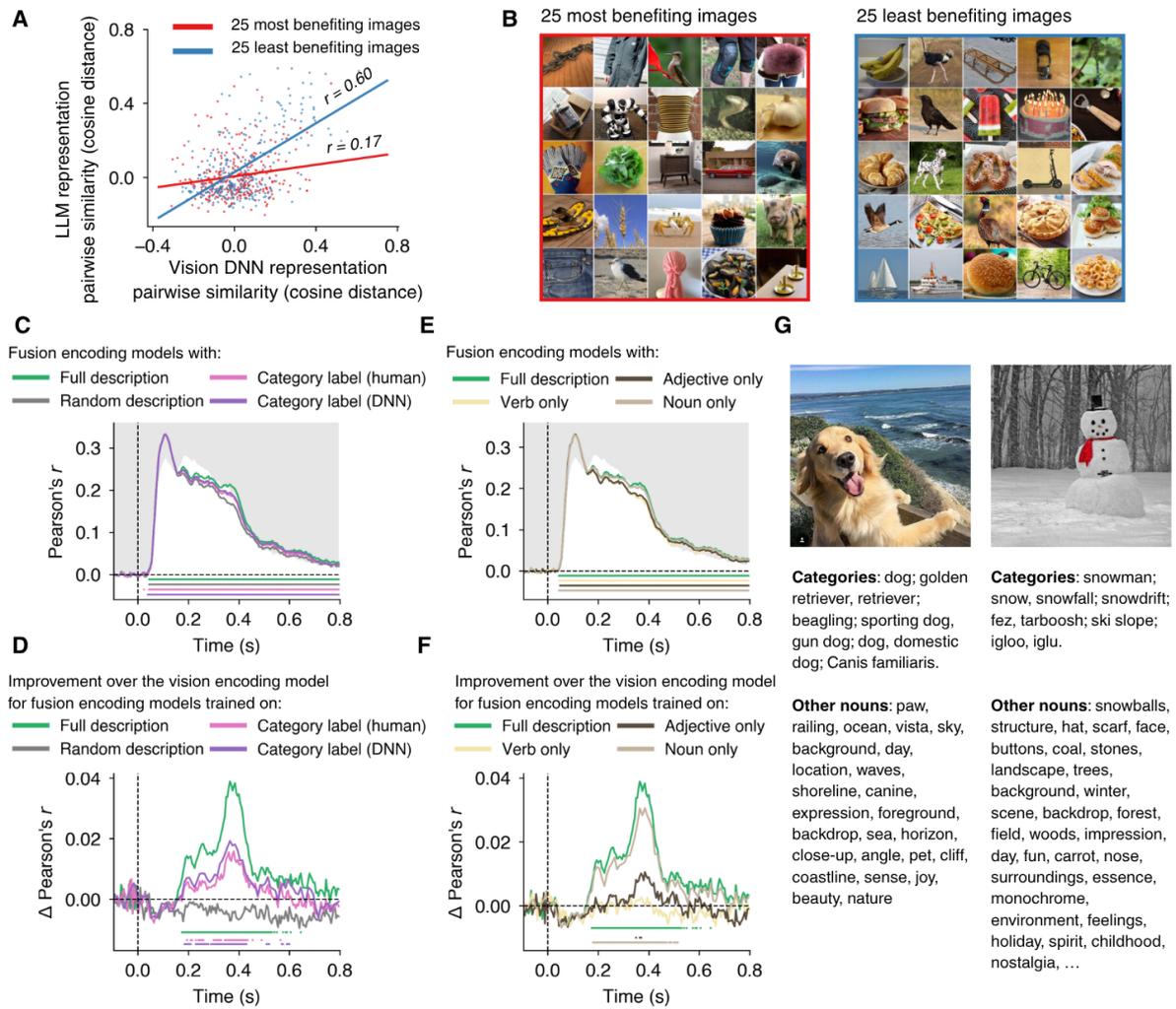

**Figure 2: Factors determining the prediction performance of the fusion encoding model. A,** Scatterplot of pairwise cosine similarity between pairs of stimulus representations of either the vision DNN or the LLM. The stimuli were the 25 stimulus images that most benefited from the fusion compared to the vision encoding model in terms of EEG prediction accuracy (color coded in red), and the 25 stimulus images that least benefitted (color coded in blue). The red and blue lines indicate the linear fit between the vision DNN and LLM pairwise similarities for the 25 most benefitting and 25 least benefitting images, respectively. **B,** Visualization of the 25 stimulus images that most or least benefited from the fusion compared to the vision encoding model in terms of EEG prediction accuracy. **C,** Prediction accuracy (Pearson's *r*) timecourse for the vision encoding model, and the fusion encoding models trained on full descriptions, object category labels, and descriptions randomly assigned to stimulus images. **D,** Prediction accuracy improvement over the vision encoding model for fusion encoding models trained on full descriptions, object category labels, and image descriptions randomly assigned to stimulus images. **E,** Prediction accuracy time course for the vision encoding model, and the fusion encoding models trained on full descriptions and on parts of speech (nouns, adjectives, and verbs from the full descriptions). **F,** Prediction accuracy improvement over the vision encoding model for fusion encoding models trained on full descriptions, nouns, adjectives, and verbs. **C-F,** The black dashed vertical lines indicate the onset of stimulus presentation, and the black dashed horizontal lines indicate the chance level of no experimental effect. Rows of asterisks at the bottom of the plots indicate significant time points (one-sided *t*-test, $p < 0.05$, FDR corrected across 180 time points, $N = 10$ participants). In gray is the area between the noise ceiling lower and upper bounds. **G,** Comparison of stimulus image object category labels (both human-annotated and DNN-generated) and nouns from the full image descriptions (excluding the nouns that overlap with the object category labels), for two illustrative examples.

## 2.2. The improvement in neural prediction of the fusion encoding model is due to differences between the vision DNN and LLM representations

Why are the LLM representations of the fusion encoding model leading to an increase in prediction accuracy compared to the vision encoding model based solely on visual DNN representations?

Above we showed that the vision DNN and the LLM differently represent stimulus information (**Fig. 1B**), suggesting that the fusion model may leverage these differences for improved neural predictivity. If so, the stimulus images that most benefit from the fusion encoding model in terms of EEG prediction accuracy, should also be most differently represented between the vision DNN and the LLM. To test this, we calculated the reduction in the mean squared error (MSE) –computed between the predicted and recorded EEG responses for each of the 200 images of the test set– for the fusion encoding model compared to the vision encoding model (i.e., $\Delta MSE = MSE_{vision} - MSE_{fusion}$). We then determined the pairwise cosine similarities of either the vision DNN or LLM representations for the group of 25 stimulus images that most benefited from the fusion model (highest $\Delta MSE$ values), and for the group of 25 stimulus images that least benefited from the fusion model (lowest $\Delta MSE$ values). Finally, we correlated the pairwise cosine similarities of the vision DNN and the LLM, independently for the stimulus pairs from each of the two stimulus groups. As expected, we observed a lower correlation for the group of stimulus images that most benefited from the fusion model (red dots, $r = 0.17$), than for the group of stimuli benefiting the least (blue dots, $r = 0.60$), with a significant difference ($\Delta r = 0.43$, $p < 10^{-10}$) (**Fig. 2A**).

Together, this indicates that the improved neural prediction performance of the fusion compared to the vision encoding model is indeed due to differences between the vision DNN and LLM representations.

Upon visual inspection, we did not observe qualitative differences between the images that most and least benefited from the fusion model (**Fig. 2B**), suggesting that the improvement in prediction accuracy is due to the language-aligned representations capturing various types of stimulus information rather than a single, qualitatively discernible, factor.

## 2.3. The LLM component of the fusion encoding model captures detailed visuo-semantic stimulus information

What type of information is captured by LLM representations so as to improve the performance of the fusion over the vision encoding model? To investigate this, we built fusion encoding models using LLM representations for different types of textual input, and compared their performance to the vision encoding model and to the fusion encoding model using LLM representations for the full image descriptions.

We first tested the relevance of object category information – a key information encoded in the brain [6,8,44,45]. In detail, we built fusion encoding models using LLM representations for the stimulus image object category annotations, generated in two ways: by humans through crowd-sourcing [46], and for the top-5 category labels generated by DNN models trained on object categorization (**Fig. 2C-D**; **Suppl. Fig. 7A**). As expected, object category information improved EEG prediction over the vision encoding model (**Fig. 2D**, pink and purple curves), but less so than the fusion encoding model based on the full image descriptions (green curve). Thus, while object category information is an important predictive factor [6,8,44,45], full descriptions contain additional visuo-semantic information predictive of neural responses.

Next, we dissected this visuo-semantic information by assessing the role of parts of speech. We reasoned that nouns reflect object-related information, adjectives reflect properties of these objects, and verbs reflect action-related information. We thus built several fusion encoding models using LLM representations for different parts of speech–nouns, adjectives, and verbs–derived from the full image descriptions (**Fig. 2E-F**; **Suppl. Fig. 7A**). We found that nouns (beige curve) and adjectives (brown curve) improved EEG prediction over the vision encoding model–although this improvement was lower than the one obtained with the full image descriptions (green curve)–and no improvement for verbs (yellow curve) (see **Suppl. Table 2** for a detailed comparison of onset and peak latency between models). This shows that the nouns and adjectives in the full image descriptions are driving the prediction accuracy improvement of the fusion over the vision encoding model, suggesting that neural responses for visual stimulation encode detailed information about objects and their properties.

The fusion encoding models trained on full image description nouns led to higher prediction accuracies compared to fusion encoding models trained on object category information (**Suppl. Fig. 7B**). To determine what additional information nouns add beyond object category labels (**Fig. 2C-D**), we compared the object category labels with the nouns from the full image description. While category labels mostly named the main object in the image, the image description nouns provided more detailed information. For example, for the image of a golden retriever, the object category labels consisted of dog breeds, while the image description nouns captured the scene context (ocean, sky, waves, shoreline), object details (paw), relations (backdrop, angle), abstract concepts (expression), and emotions (joy) (**Fig. 2G**, left). Similarly, for the image of a snowman, the image description nouns provided detailed information about object parts (scarf, buttons, coal, carrot), background elements (landscape, trees, forest, winter), and associated emotional and abstract concepts (fun, childhood, nostalgia), that were absent from the object category labels (**Fig. 2G**, right).

Finally, as a control we ascertained that our results are not trivially driven by adding LLM representations per se, independent of their meaning. For this we built a fusion encoding model based on LLM representations for full image descriptions that were randomly assigned to images. This led to no significant improvement in prediction accuracy compared to the vision encoding model (**Fig. 2C-D**, gray curve). Thus, the improvement in prediction accuracy of the fusion encoding model (with image descriptions correctly assigned to stimulus images) over the vision encoding model is driven by meaningful information contained in the image descriptions.

Together, this shows that LLM representations capture detailed visuo-semantic information about the stimulus images, accounting for the improvement in prediction accuracy of the fusion over the vision encoding model.

## 2.4. The fusion encoding model outperforms multimodal models and is robust to model choice

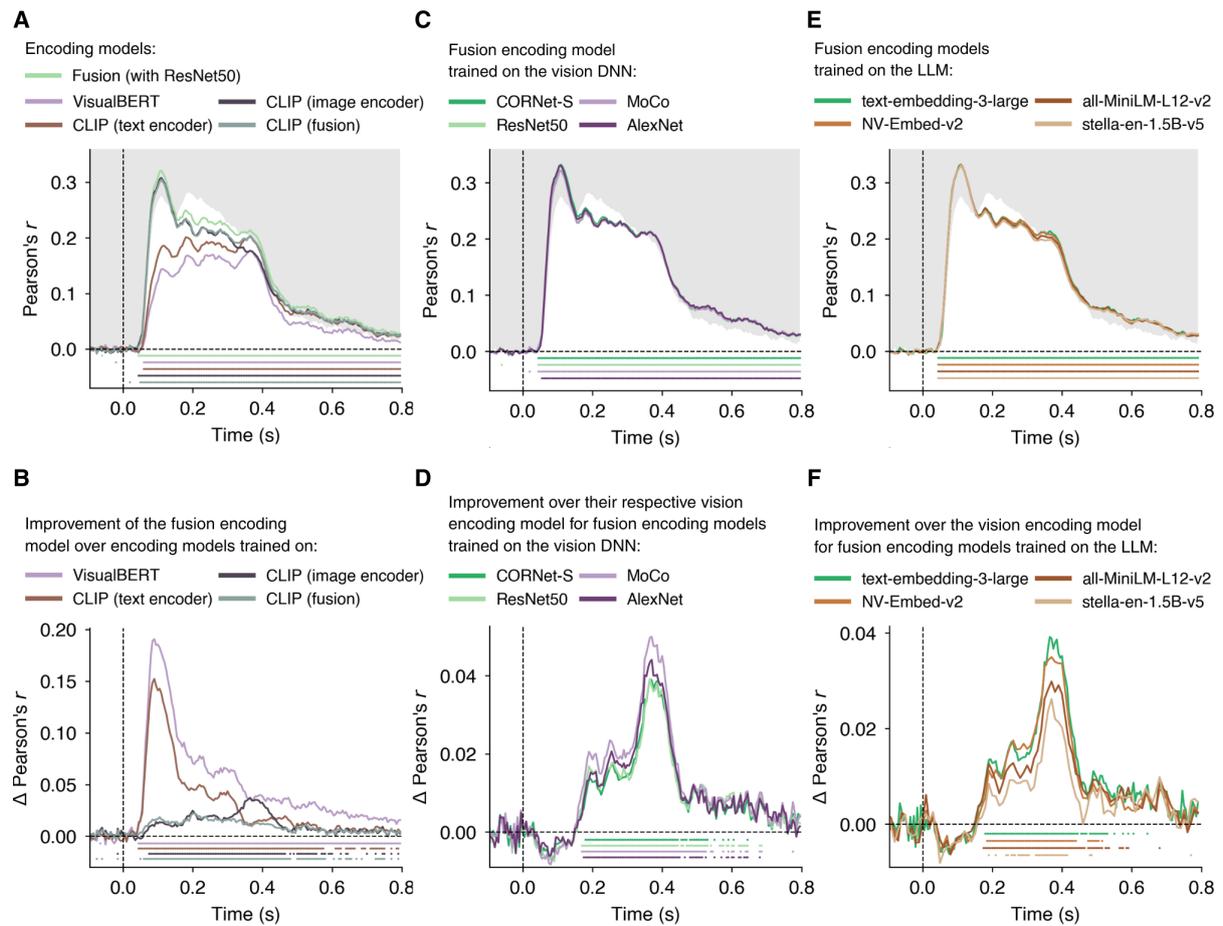

**Figure 3: Model comparisons and generalizability analysis. A,** Prediction accuracy timecourse for the fusion encoding model, and the encoding models trained on representations from multimodal DNNs (CLIP and VisualBERT). **B,** Difference in prediction accuracy between the fusion encoding model, and the encoding models trained on representations from multimodal DNNs (CLIP and VisualBERT). **C,** Prediction accuracy timecourse for fusion encoding models trained using different vision DNNs. **D,** Prediction accuracy improvement over the vision encoding model for fusion encoding models trained using different vision DNNs. **E,** Prediction accuracy timecourse for fusion encoding models trained using different LLMs. **F,** Prediction accuracy improvement over the vision encoding model for fusion encoding models trained using different LLMs. **A-F,** The black dashed vertical lines indicate the onset of stimulus presentation, and the black dashed horizontal lines indicate the chance level of no experimental effect. Rows of asterisks at the bottom of the plots indicate significant time points (one-sided $t$-test, $p < 0.05$, FDR corrected across 180 time points, $N = 10$ participants). In gray is the area between the noise ceiling lower and upper bounds.

Previous research established that multimodal DNNs trained on combined visual and linguistic input result in more accurate encoding models compared to vision encoding models, especially beyond the first stages of neural responses to visual stimulation [21,31,47–50]. We thus benchmarked our fusion encoding model against encoding models built using two multimodal DNNs: CLIP [25,29,51] and VisualBERT [24,52]. We found that our fusion encoding model outperformed both multimodal models (**Fig. 3A-B**), demonstrating the advantage of the fusion model approach for neural prediction.

Finally, to ascertain that our results do not depend on the exact instantiation of the vision DNN and LLM used, we built fusion encoding models using different vision DNNs and LLMs (**Fig. 3C-E**). Comparing the prediction accuracy of these alternative fusion encoding model instantiations against the prediction accuracy of the respective vision encoding models revealed a significant improvement in all cases, demonstrating the generality of our findings (**Fig. 3D,F**; see **Suppl. Table 2** for a detailed comparison of onset and peak latency between models).

## 2.5. The vision DNN and LLM capture distinct spectral signatures of neural activity

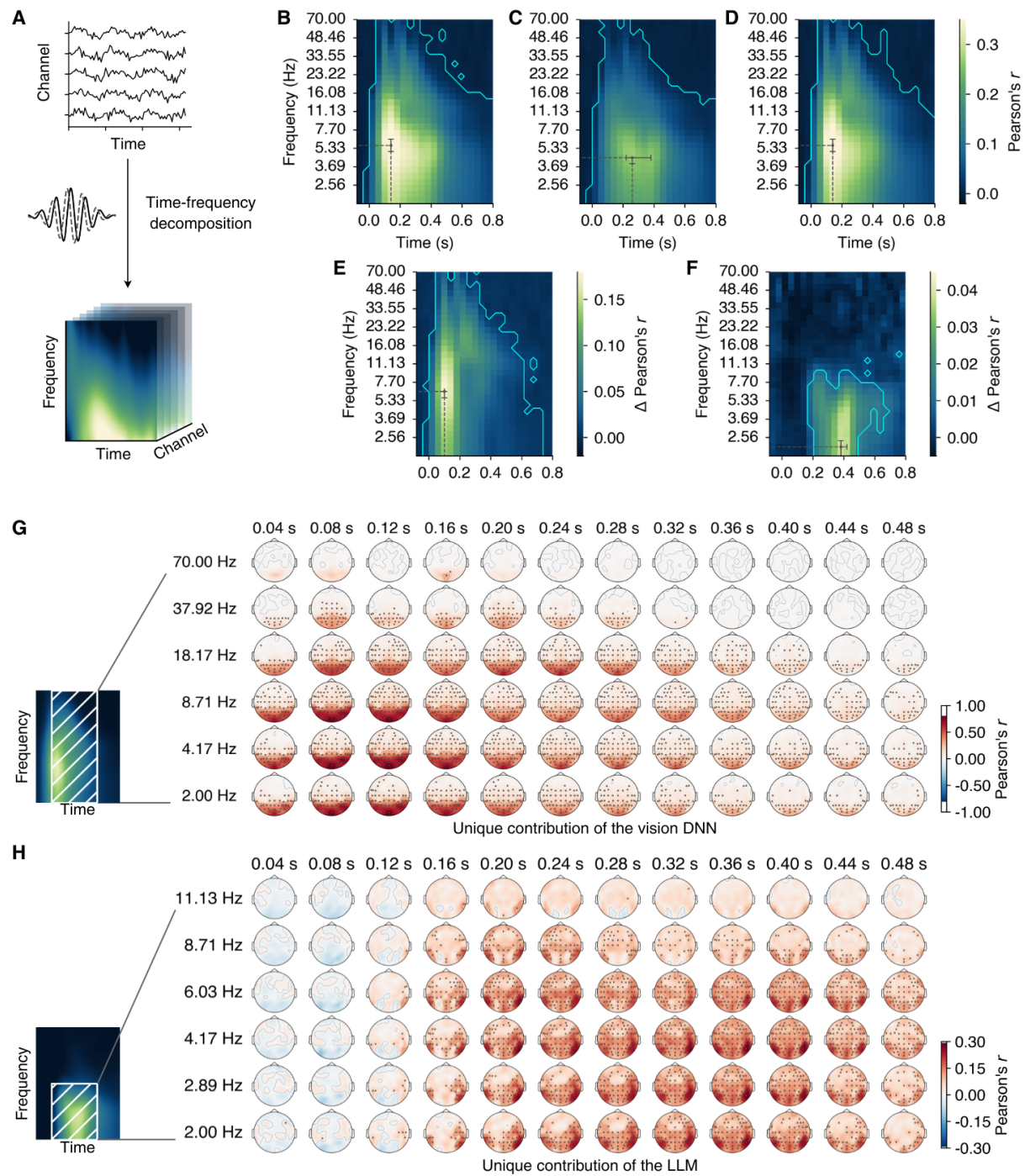

**Figure 4: Time-frequency resolved analysis and results. A,** We decomposed the EEG responses into the time-frequency domain using Morlet wavelets. **B-D,** Prediction accuracy (Pearson's *r*) of the EEG time-frequency data

of the vision (**B**), language (**C**), and fusion (**D**) encoding models. **E,** Difference in prediction accuracy between the fusion and the language encoding models, which isolated the effect of the vision DNN. **F,** Difference in prediction accuracy between the fusion and the vision encoding models, which isolated the effect of the LLM.. **B-F,** The prediction accuracies are averaged across all participants and EEG channels. The gray dashed lines indicate latency and frequency peaks of prediction accuracy with 95% confidence intervals. Cyan contour lines delineate clusters of significant effects (one-sided *t*-test, $p < 0.05$, FDR-corrected across 30 frequency points and 180 time points, $N = 10$ participants). **G,** EEG topography of partial correlation results, indicating the unique prediction accuracy contribution of the vision DNN. **H,** EEG topography of partial correlation results, indicating the unique prediction accuracy contribution of the LLM. **G-H,** The highlighted black dots indicate significant channels (one-sided *t*-test, $p < 0.05$, FDR-corrected across 63 channels and 180 time points, $N = 10$ participants).

Our results suggested that the contributions of the vision DNN and LLM to the fusion encoding model depend on distinct cortical circuits active at different time points. Based on this, we hypothesized that the vision DNN and the LLM also differently relate to the spectral basis of the EEG response [53–55]. Specifically, since early neural responses with respect to image onset relate to broadband neural activity [56–61], we hypothesized that the vision DNN component of the fusion model would accurately predict the full frequency spectrum of EEG responses. Furthermore, since later neural responses relate to low-frequency neural activity [62–64], we further hypothesized that the LLM component of the fusion model would accurately predict low-frequencies of EEG responses.

To test this, we decomposed the EEG signals into time-frequency representations from 2 to 70 Hz (**Fig. 4A**), and trained encoding models to predict the EEG responses in this time-frequency domain.

We found that the vision (**Fig. 4B**), language (**Fig. 4C**), and fusion (**Fig. 4D**) encoding models predicted neural activity across all investigated frequencies (cyan contours delineate significant effects. To directly assess differences between models, we subtracted the prediction accuracy of the vision (**Fig. 4E**) or language (**Fig. 4F**) encoding models from the prediction accuracy of the fusion encoding model . The difference between the fusion and the language encoding models, which isolated the effect of the vision DNN, peaked at 80 ms (80 – 80 ms), with broadband effect across the whole investigated frequency range peaking at 6.82 Hz (6.82 – 7.70 Hz) (**Fig. 4E**). In contrast, the difference between the fusion and the vision encoding models, which isolated the effect of the LLM, peaked at 360 ms (360 – 400 ms), with effects limited to frequencies below 12Hz peaking at 2.26 Hz (2.00 – 4.72 Hz) (**Fig. 4F**). These peaks had a significant latency difference of 280 ms (280 – 320 ms, $p < 10^{-5}$), and a significant frequency difference of 4.55 Hz (2.10 – 5.70 Hz, $p < 10^{-5}$) (for statistical details see **Suppl. Table 3**). Together, this further confirms that the vision DNN and LLM components uniquely capture earlier or later stages of EEG responses to visual stimuli, respectively, and additionally reveals that these two components capture different spectral bases of the EEG response. We obtained an equivalent results pattern when using partial correlation (**Suppl. Fig. 8**) and variance partitioning (**Suppl. Fig. 9**) instead of subtraction, thus substantiating the results.

To estimate the cortical sources uniquely predicted by the vision DNN or the LLM, we partialled out the contribution of the language or vision encoding model on the fusion encoding model, respectively, and inspected the resulting partial correlation topographies in EEG sensor space (**Fig. 4G-H**) ($N = 10$ participants, one-sided *t*-test, $p < 0.05$, FDR-corrected across 63 channels and 180 time points). The earlier and broadband frequency effects (from 2 Hz to 70 Hz) for the vision model peaked at occipital electrodes (e.g., Oz, POz), whereas the later and lower frequency effects (from 2 Hz to 12 Hz) for the language model peaked in bilateral occipito-temporal electrodes (e.g., TP9, TP10), suggesting occipital and temporal cortical regions as sources of broad- and low-frequency bands, respectively (for statistical details see **Suppl. Table 3**, for visualization across the full frequency range see **Suppl. Fig. 10-11**). Equivalent results pattern emerged using variance partitioning (**Suppl. Fig. 12-13**) and subtraction (**Suppl. Fig. 14-15**).

Together, this revealed that the vision DNN and LLM components of the fusion encoding model capture neural dynamics with distinct spectral signatures and cortical origin: the vision DNN component captured early, broadband responses from occipital electrodes, whereas the LLM component captured later, low-frequency responses from occipito-temporal electrodes.

# 3. Discussion

Our goal was to provide a modelling approach accounting for the time course of visuo-semantic processing in the human brain. Towards this goal, we trained encoding models on a large datasets of EEG responses to naturalistic images to test the hypothesis that human representations across the visual processing hierarchy are better modelled by a fusion of representations from a vision DNN and an LLM, than by the representations of either model alone. Our results confirmed this hypothesis, and also showed that the fusion encoding model outperformed previous multimodal approaches. This predictive benefit was due to the vision DNN and LLM components capturing complementary neural representations: the vision DNN captured earlier and broadband signals, whereas the LLM captured later and low frequency signals. Investigation into the factors responsible for the prediction improvement by the LLM component revealed a role for detailed visuo-semantic stimulus information beyond the category of the objects viewed. The overall results pattern was robust across different instantiations of vision DNNs and LLMs, demonstrating the generalizability of our approach.

## 3.1. Combining a vision DNN with an LLM improves the prediction of neural responses to visual stimulation

Building an encoding model of human visual processing is an integral part in the quest of understanding human vision [5,9,18,65]. While, in the last decade, the modelling of human visual processing has been dominated by vision DNNs [9–11,13,14,19], recent research started exploring language models that capture visuo-semantic information – both LLMs [21,49,66] and multimodal DNNs that combine visual and linguistic information [21,29,47,48,51,52,67]. Here we showed that fusing a vision DNN and an LLM outperformed encoding models trained on the two unimodal components in isolation in predicting human EEG responses to visual stimulation. The improvement over the unimodal encoding models lies in fusing the predictive power of vision DNNs and LLMs in two ways. First, the fusion encoding model reaches the same prediction accuracy as the vision or language encoding models at their respective peak latencies (i.e., 105 ms and 365 ms, respectively). Second, during the interim processing period between peaks, the fusion encoding model outperforms vision and language encoding models. Thus, the fusion encoding model predicts the temporal dynamics of neural responses to visual stimuli either equally well, or better, than vision or language encoding models.

Our fusion encoding model also outperformed previous approaches focussing on multimodal architectures, in particular VisualBERT [24,52] and CLIP [25,29,51]. This offers exciting venues for future research. CLIP, VisualBERT, and similar multimodal architectures provide state-of-the-art applications in a wide range of cognitive scenarios such as neural predictions with audiovisual stimuli [49,67], cross-species neural predictions [66,68], and prediction of behavioral patterns [22,23]. Given that our fusion encoding model outperformed these multimodal approaches, this invites direct extension of the fusion model to the aforementioned applications for further increases in neural prediction.

Rather than providing one, unified model of neural processing, our fusion encoding model divides and conquers the neural prediction challenge in two parts, carried out by the vision DNN and LLM components, respectively. This allows to quantify how much do vision DNN and LLM representations contribute to neural prediction, in contrast to multimodal architectures where these representations are highly entangled [69]. We leveraged this interpretability to determine the factors determining the added predictive benefit of the LLM representations over the vision DNN representations. We found that these factors consist of detailed visuo-semantic stimulus information, including context, object parts, material properties, relations, and emotional connotations. This supports the idea that the objective of the ventral visual stream goes beyond the categorization of the major objects present in a scene [6,8,44,45], instead including a detailed semantic analysis of visual scenes [70–73].

### 3.2. The vision DNN and LLM characterize distinct neural processing stages

The vision DNN and LLM representations revealed distinct neural spectro-temporal signatures. As expected, the vision DNN uniquely accounted for early stages of visual processing (i.e., with a prediction accuracy peak at 110 ms over occipital electrodes), and for a broadband spectral basis (2 – 70 Hz). This timing is consistent with an early processing stage in hierarchical models of vision [5,8], and matches the spectral signatures of processing in early visual cortex [55,74,75].

In contrast, the LLM uniquely accounted for later stages of neural processing, revealing [31] two processing stages mostly in the delta (2 – 4 Hz) and theta (4 – 8 Hz) frequency range defined by two peaks: an earlier peak at 200 ms overlying temporal cortex, and a later peak at 365 ms.

We hypothesize that the first peak at 200 ms coincides with the first emergence of conceptual object representations [88], suggestive of early semantic processing driven by theta oscillations in the anterior temporal lobe (ATL) [77–80]. According to the hub-and-spoke theory [81–83] of semantic representation, the ATL serves as a transmodal semantic hub that rapidly activates conceptual knowledge based on incoming low-level information.

The second peak at 365 ms had a more widespread topography, including electrodes overlying the frontal cortex. In both timing and topography, this second peak parallels the N400 component [84,85], which indexes semantic memory access and integration through an extensive frontotemporal network [86,87] linked to the theta frequency band [88–91].

### 3.3. Conclusion

Explaining human vision involves providing an encoding model of the neural representations that mediate our rich and detailed perception of the world. Our results indicate that combining vision DNNs with LLMs yields encoding models that outperform previous approaches in the prediction accuracy of time-resolved human EEG responses to visual stimuli. This opens a new direction for modelling and characterizing the time course of visuo-semantic processing in the human brain.

# 4. Methods

## 4.1. Data

For all our analyses we used the THINGS EEG2 dataset [32], a large-scale dataset of EEG responses to naturalistic images optimized to build encoding models of neural responses to visual stimulation. The THINGS EEG2 dataset contains EEG recordings of 10 participants viewing images of objects on natural backgrounds from the THINGS database [46]. Each participant completed 82,160 trials, over 4 data collection sessions, spanning 16,740 distinct images. The dataset is split into a training partition of 16,540 images each presented 4 times (i.e., 4 repetitions per image), and a test image partition of 200 images each presented 80 times (i.e., 80 repetitions per image). The images in the training partition belong to 1,654 object categories (i.e., 10 exemplar images per category), whereas the images in the test partition belong to 200 object categories that do not overlap with the categories from the training partition.

The EEG recordings were collected using a 64-channel EASYCAP with electrodes arranged in the standard 10–10 system [92], and a Brainvision actiCHamp amplifier at a sampling rate of 1,000 Hz. The data were online filtered between 0.03 and 100 Hz and referenced to the Fz electrode.

### 4.1.1. EEG preprocessing

We epoched the EEG trials from -100 ms to 800 ms with respect to stimulus onset. To sharpen the spatial resolution of the EEG signal, we applied a current source density (CSD) transform to the epoched data. The CSD transform computes the spatial Laplacian of the sensor signal, which reduces volume conduction effects and minimizes the spatial blurring of the EEG responses [93–95]. Next, we downsampled the data to 200 Hz and baseline-corrected it by subtracting the mean of the pre-stimulus interval for each trial and channel separately. Finally, to correct for session-specific effects we *z*-scored the data within each recording session.

### 4.1.2. EEG time-frequency decomposition

For the EEG time-frequency domain analysis, we computed the time-frequency decomposition on the EEG dataset using complex Morlet wavelets [96,97], resulting in a complex vector at each selected time-frequency point and channel. The frequencies ranged from 2 to 70 Hz in 50 logarithmically spaced increments. The time intervals ranged from -80 ms to 800 ms with respect to stimulus onset, in steps of 40 ms. To ensure a uniform temporal smoothing, we set a fixed number of cycles (i.e., 5 cycles) across all frequencies.

## 4.4. Encoding models

We trained encoding models of EEG responses to images using image feature representations from vision DNNs, LLMs, and their combination.

### 4.4.1. Vision DNN image feature representations

As the vision DNN we used CORnet-S [40], a recurrent convolutional neural network composed of four sequential, anatomically-mapped areas (V1, V2, V4 and IT) with within-area recurrence, trained on object categorization on ImageNet [98].

For the main analyses, we extracted the image feature representations of CORnet-S from the last layers of areas V1, V2, V4, IT, and from the decoder layer. We then $z$-scored the feature maps of the training partition images to zero mean and unit variance for each feature across the image sample dimension, and applied nonlinear principal component analysis (PCA) (polynomial kernel of degree 4) to reduce the features to 1,000 components. Finally, we applied the same $z$-score and PCA steps to the test partition images using the mean, standard deviation, and PCA weights fit with the training partition images.

For the control analyses, we built encoding models using image feature representations extracted from three additional vision DNNs: (1) AlexNet [99], a supervised feedforward DNN with 5 convolutional layers followed by 3 fully-connected layers, from which we extracted feature maps from layers maxpool1, maxpool2, ReLU3, ReLU4, maxpool5, ReLU6, ReLU7, and fc8; (2) ResNet-50 [100], a supervised model of 50 layers incorporating residual connections, from which we extracted feature maps from the last layer of each of its four main blocks and from the decoder layer; and (3) MoCo [101], which consists in the same architecture as ResNet-50 trained in a self-supervised manner, from which we extracted feature maps from the same layers as ResNet-50. We applied the same $z$-score and PCA steps described above to the image feature maps of these additional models.

### 4.4.2. Image description generation and LLM feature representations

We generated five distinct text descriptions for each image in the dataset, by feeding the images to GPT-4V [102] with the prompt: *"Describe the image in five different ways"*. Because 45 train images did not pass the safety check of GPT-4V, we excluded those images and trained the models with the remaining 16,495 image conditions. The generated descriptions formed the input to the LLMs.

For the main analyses, we extracted the LLM feature representations using OpenAI's text-embedding-3-large. Specifically, we fed the image descriptions to text-embedding-3-large, averaged the resulting embeddings across the five description versions (i.e., resulting in one LLM feature representation for each image), and applied the same $z$-score and PCA steps described above to the LLM feature representations.

For the control analyses, we built encoding models using LLM feature representations extracted from three additional LLMs from the MTEB leaderboard [103]: NV-Embed-v2, all-MiniLM-L12-v2 and stella-en-1.5B-v5. We applied the same $z$-score and PCA steps described above to the LLM feature representations of these additional models.

Finally, to determine the type of semantic information relevant for predicting neural activity, we passed the following types of text into the OpenAI's text-embedding-3-large model: human-annotated object categories, DNN-generated object categories, permuted descriptions, nouns from the descriptions, adjectives from the descriptions, and verbs from the descriptions.

### 4.4.3. Encoding model training

The encoding models consisted in ridge regressions that mapped vision DNN and/or LLM representations onto EEG responses. Using the training image partitions we trained independent ridge regression for each subject, EEG channel, and EEG time point. We started by averaging the EEG responses for each training image across the 4 repetitions. Next, we trained three distinct encoding models:

1. A vision encoding model, trained using the vision DNN representations as predictosrs.

2. A language encoding model, trained using LLM representations as predictors.

3. A fusion encoding model, trained using a combination of vision DNN and LLM representations as predictors. To combine the vision DNN representations (X) with the LLM representations (Y), we used a convex combination. Specifically, we computed the fused representations as a concatenation of weighted X and Y: $[(1-\alpha)X, \alpha Y]$, where $\alpha$ is a weighting coefficient constrained to $0 \leq \alpha \leq 1$.

For model training, we used a grid search algorithm [104] with 5-fold cross-validation to optimize the ridge regularization strength $\lambda$ (in the range $[10^{-3}, 10^{4}]$ with 100 log-spaced steps), and the weighting coefficient $\alpha$ for the fusion model (in the range $[0, 1]$ with a step size of 0.05), independently for each subject, EEG channel, and EEG time point.

### 4.4.4. Encoding model testing

We tested the encoding models' performance by comparing the predicted EEG responses for the test images to the recorded EEG responses for the same test images. In detail, we first randomly selected and averaged 40 repetitions of the recorded EEG responses for each test image (we used the other 40 repetitions to estimate the noise ceiling, see "Noise ceiling calculation" section). Then, for each subject, EEG channel, and EEG time point, we calculated the Pearson's correlation coefficient between the 200-dimensional vectors of the predicted and averaged recorded EEG responses for the 200 test images. We iterated this process 100 times, each time averaging across a different random split of recorded EEG data repetitions, and averaged the correlation coefficients across the 100 iterations.

### 4.4.5. Noise ceiling calculation

We calculated the noise ceiling lower and upper bounds to estimate the theoretical maximal encoding predictions performance given the noise in the recorded EEG data. We randomly divided the recorded EEG responses for the test images into two non-overlapping partitions of 40 image condition repetitions each, where the first partition corresponded to the 40 repetitions of recorded EEG responses described in section "4.4.4 Encoding model testing". To obtain the noise ceiling lower bound estimate, we averaged the recorded EEG responses of each partition across repetitions, and correlated the resulting averaged responses between the two partitions. To obtain the noise ceiling upper bound estimate, we correlated the recorded EEG responses of the first partition averaged across the 40 repetitions, with the recorded EEG responses averaged across all 80 test image repetitions. To avoid the results being biased by one specific configuration of repetitions, we iterated the correlation analyses 100 times, while always randomly assigning different repetitions to the two recorded EEG response partitions data partitions, and then averaged the results across the 100 iterations.

### 4.4.6. Partial correlation analysis

We performed a partial correlation analysis to isolate the individual contribution of vision DNNs and LLMs to the prediction accuracy of the fusion model. We conducted two types of analyses for each subject, EEG channel, and EEG time point:

1. To determine the individual contribution of the vision DNN, we calculated the partial correlation between recorded EEG responses and EEG responses predicted by the fusion encoding model, while partialling out (i.e., controlling) the EEG variance explained by the language encoding model.
2. To determine the individual contribution of the LLM, we calculated the partial correlation between recorded EEG responses and EEG responses predicted by the fusion encoding model, while partialling out the EEG variance explained by the vision encoding model.

In each case the partial correlation analysis consisted in the following steps:

1. Using ordinary least squares (OLS) we regressed the variance of the recorded EEG responses and of the EEG responses predicted by the fusion encoding model, that is explained by the control encoding model (i.e., the vision or language encoding model).
2. We computed the residuals of both regressions in 1.
3. We computed the Pearson's correlation coefficients between the two sets of residuals.

Here too we repeated the analysis across 100 iterations, and averaged the results across iterations, as described above.

### 4.4.7. Encoding models training and testing for time-frequency EEG data

To train and test encoding models of the EEG responses in the time-frequency domain we followed the same analysis rationale as described above, with the following two differences. First, we additionally trained and tested a separate encoding model for each frequency. Second, as the time-frequency decomposed EEG responses are complex values with a real part and an imaginary part, for all analysis steps involving correlations (i.e., during encoding model testing) we used complex Pearson correlation coefficients [105,106].

### 4.5. Statistical testing

To establish the statistical significance of the correlation, decoding, and variance partitioning analyses, we tested all results against chance using one-sided *t*-tests. The rationale was to reject the null hypothesis H0 of the analyses results being at chance level with a confidence of 95% or higher (i.e., with a *p* value of $p < 0.05$), thus supporting the experimental hypothesis H1 of the results being significantly higher than chance. The chance level differed across analyses: 0 Pearson's *r* in the correlation analyses; 0 % accuracy in the pairwise decoding analyses; 0 explained variance in the variance partitioning analysis; a difference of 0 in the comparisons between the different encoding models.

We controlled familywise error rate by applying a false discovery rate (FDR) correction [107] to the resulting P-values to correct for the number of EEG time points ($N = 180$) in all analyses, and additionally for the number of EEG channels ($N = 63$) in the EEG topoplot visualizations.

To calculate the 95% confidence intervals of the peak latencies and peak latency differences, we created 10,000 bootstrapped samples by resampling the subject-specific results with replacement. This yielded empirical distributions of the results, from which we took the 95% confidence intervals.

## Data availability

Our analyses were based on THINGS EEG2, a freely available dataset of EEG recordings of 10 participants viewing images of 16,740 objects on natural backgrounds 32. THINGS EEG2's raw EEG data is available on OSF at https://osf.io/3jk45/ and the preprocessed EEG data is available on OSF at https://osf.io/3eayd/.

## Code availability

Our code is available as a public Github repository at https://github.com/rbybryan/EEG_fusion_encoding.

## Acknowledgments

R.M.C. is supported by German Research Council (DFG) grants (CI 241/1-3, CI 241/1-7, INST 272/297-2), European Research Council (ERC) Consolidator Grant (ERC-CoG-2024101123101). We thank the HPC Service of FUB-IT, Freie Universität Berlin, for computing time (DOI: http://dx.doi.org/10.17169/refubium-26754).

## Author contributions

E.D. and R.M.C. acquired funding. B.R. and R.M.C. designed research. B.R. and A.T.G. preprocessed EEG data. B.R. modelled and analyzed data. B.R., A.T.G. and R.M.C. interpreted results. B.R. prepared figures. B.R. drafted the manuscript. B.R., A.T.G., E.D. and R.M.C. edited and revised the manuscript. All authors approved the final version of the manuscript.

## Competing interests

The authors declare no competing interests.

30. Tang, J., Du, M., Vo, V. A., Lal, V. & Huth, A. G. Brain encoding models based on multimodal transformers can transfer across language and vision. *Adv. Neural Inf. Process. Syst.* **36**, 29654–29666 (2023).

31. Grosbard, I. D. & Yovel, G. Self-supervision deep learning models are better models of human high-level visual cortex: The roles of multi-modality and dataset training size. 2025.01.09.632216 Preprint at https://doi.org/10.1101/2025.01.09.632216 (2025).

32. Gifford, A. T., Dwivedi, K., Roig, G. & Cichy, R. M. A large and rich EEG dataset for modeling human visual object recognition. *NeuroImage* **264**, 119754 (2022).

33. Kay, K. N., Naselaris, T., Prenger, R. J. & Gallant, J. L. Identifying natural images from human brain activity. *Nature* **452**, 352–355 (2008).

34. Naselaris, T., Kay, K. N., Nishimoto, S. & Gallant, J. L. Encoding and decoding in fMRI. *NeuroImage* **56**, 400–410 (2011).

35. Rafegas, I. & Vanrell, M. Color Representation in CNNs: Parallelisms with Biological Vision. *2017 IEEE Int. Conf. Comput. Vis. Workshop ICCVW* 2697–2705 (2017) doi:10.1109/ICCVW.2017.318.

36. Baker, N., Lu, H., Erlikhman, G. & Kellman, P. J. Local features and global shape information in object classification by deep convolutional neural networks. *Vision Res.* **172**, 46–61 (2020).

37. Zeman, A. A., Ritchie, J. B., Bracci, S. & Op de Beeck, H. Orthogonal Representations of Object Shape and Category in Deep Convolutional Neural Networks and Human Visual Cortex. *Sci. Rep.* **10**, 2453 (2020).

38. Piantadosi, S. T. & Hill, F. Meaning without reference in large language models. Preprint at https://doi.org/10.48550/arXiv.2208.02957 (2022).
24

https://doi.org/10.1101/2025.03.05.641284 (2025).

67. Small, H., Lee Masson, H. & Isik, L. Vision and language representations in multimodal AI models and human social brain regions during natural movie viewing. In *UniReps: Unifying Representations in Neural Models Workshop*, 38th Conference on Neural Information Processing Systems (NeurIPS), 2024.

68. Conwell, C. *et al*. Is visual cortex really "language-aligned"? Perspectives from Model-to-Brain Comparisons in Human and Monkeys on the Natural Scenes Dataset. *J. Vis*. **24**, 1288 (2024).

69. Dang, Y. *et al*. Explainable and Interpretable Multimodal Large Language Models: A Comprehensive Survey. Preprint at https://doi.org/10.48550/arXiv.2412.02104 (2024).

70. Russell, B., Torralba, A., Liu, C., Fergus, R. & Freeman, W. Object Recognition by Scene Alignment. in *Advances in Neural Information Processing Systems* vol. 20 (Curran Associates, Inc., 2007).

71. Brandman, T. & Peelen, M. V. Interaction between Scene and Object Processing Revealed by Human fMRI and MEG Decoding. *J. Neurosci*. **37**, 7700–7710 (2017).

72. Võ, M. L.-H., Boettcher, S. E. & Draschkow, D. Reading scenes: how scene grammar guides attention and aids perception in real-world environments. *Curr. Opin. Psychol*. **29**, 205–210 (2019).

73. Võ, M. L.-H. The meaning and structure of scenes. *Vision Res*. **181**, 10–20 (2021).

74. Buffalo, E. A., Fries, P., Landman, R., Buschman, T. J. & Desimone, R. Laminar differences in gamma and alpha coherence in the ventral stream. *Proc. Natl. Acad. Sci*. **108**, 11262–11267 (2011).

75. Pantazis, D. *et al*. Decoding the orientation of contrast edges from MEG evoked and

# Supplementary Figures

**Supplementary Figure 1: Encoding models' single participant prediction accuracy (correlation)**

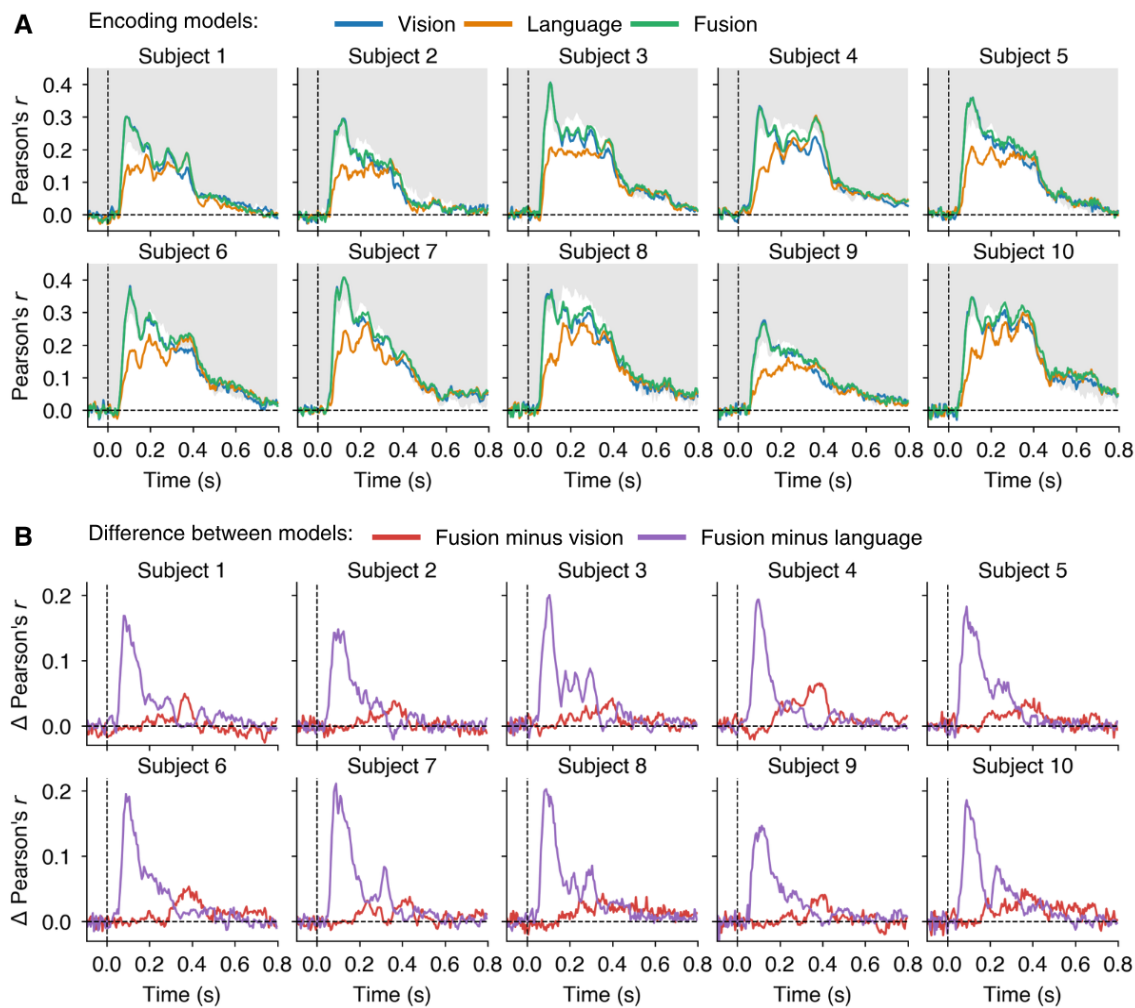

**A,** Prediction accuracy (Pearson's *r*) correlations time course for the visual, language, and fusion encoding models on single-subject level. In gray is the area between the noise ceiling lower and upper bounds. **B,** Difference in prediction accuracy between the fusion and the vision or language encoding model on single-subject level. **A-B,** The black dashed vertical lines indicate the onset of stimulus presentation, and the black dashed horizontal lines indicate the chance level of no experimental effect. In gray is the area between the noise ceiling lower and upper bounds.

# Supplementary Figure 2: Encoding models' prediction accuracy (pairwise decoding)

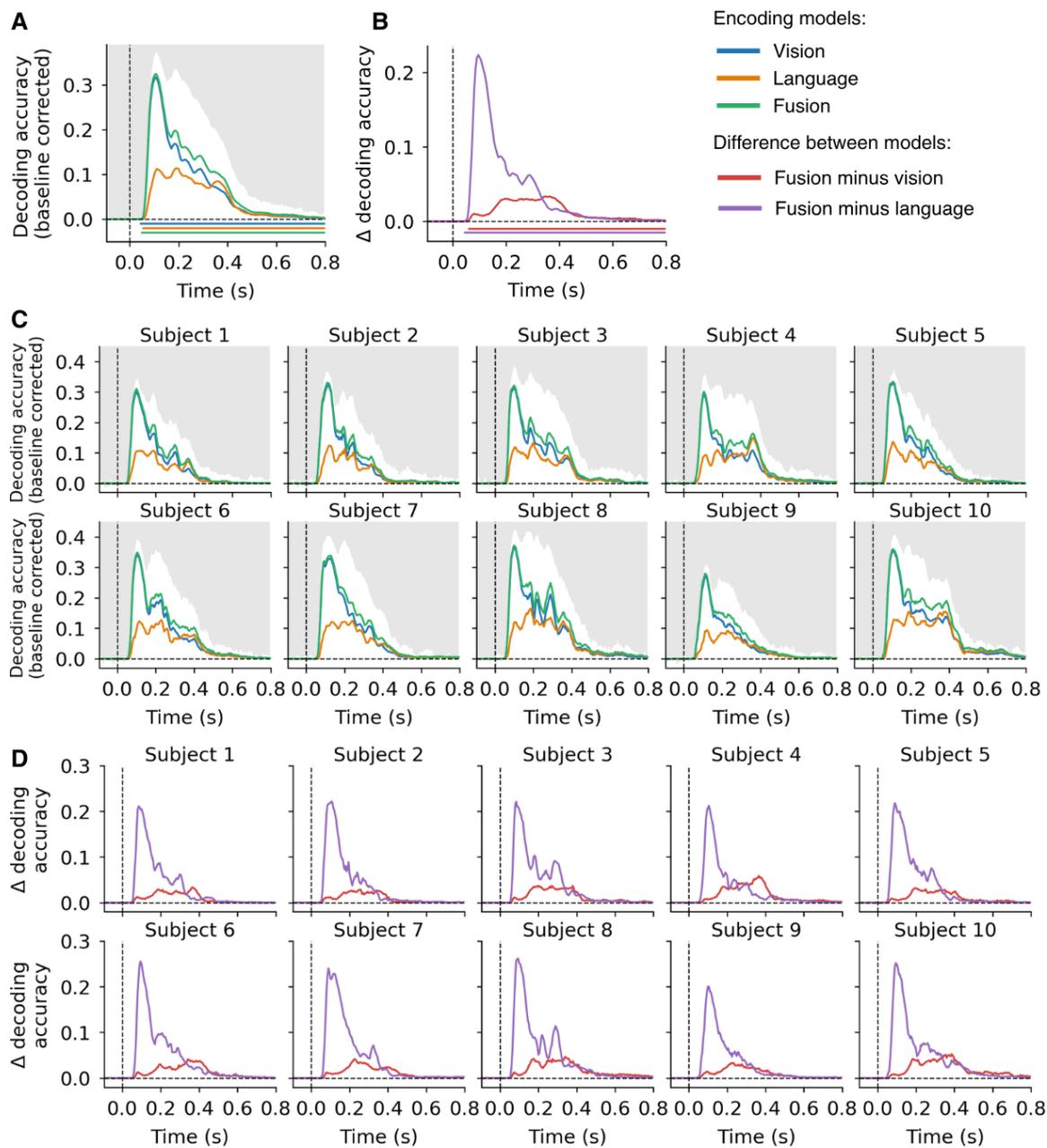

The rationale of this analysis was to see if a classifier trained on the recorded EEG responses is capable of generalizing its performance to the predicted EEG responses from the encoding models. This is a complementary way (to the correlation analysis, see **Fig. 1E-F, Suppl. Fig. 1**) to assess the similarity between the recorded and predicted EEG responses, hence the encoding models' predictive power. We performed the pairwise decoding analysis using the recorded and predicted EEG responses for the 200 test images. We started by averaging 40 recorded EEG image condition repetitions (we used the other 40 repetitions to estimate the noise ceiling; see the "Noise ceiling calculation" paragraph in the Methods section) into 10 pseudo-trials of 4 repeats each. Next, we used the pseudo-trials for training linear support vector machines (SVMs) to perform binary classification between each pair of the 200 recorded EEG image conditions using their EEG channel activity. We then tested the trained classifiers on the corresponding pairs of predicted EEG image conditions. We performed the pairwise decoding analysis independently for each EEG time point, and then averaged decoding accuracy scores across image condition pairs, obtaining a time course of decoding accuracies. **A,** Pairwise decoding accuracy time course between the recorded and predicted EEG responses from visual, language, and fusion encoding models. The decoding accuracy using the vision encoding model peaks at 105 ms (100-110 ms); the decoding accuracy using the language encoding model peaks 190 ms (110-200 ms); the decoding accuracy using the fusion encoding model peaks 105 ms (100-110 ms). There is no peak-to-peak latency difference between the fusion encoding model and vision encoding model (0 ms (0-5 ms), $p > 0.05$). The peak-to-peak latency difference between the fusion and language encoding models is 85 ms (0-95 ms), $p = 0.06$). **B,** Difference in decoding accuracy between the fusion encoding model and the vision or language encoding models. The difference between the fusion and vision encoding models peaks at 365 ms (210-365 ms); the difference between the fusion and language encoding models peaks at 95 ms (90-100 ms). Their peak-to-peak latency difference is 270 ms (115-270 ms, $p < 10^{-4}$). **A-B,** Colored dots below plots indicate significant points (one sided t-test, p < 0.05, FDR-corrected for 180 time points, N = 10 participants). **C,** Decoding accuracy time course for visual, language, and fusion encoding models on single-subject level. **D,** Difference in decoding accuracy between the fusion encoding model and the vision or language encoding models on single-subject level. **A and C,** In gray is the area between the noise ceiling lower and upper bounds. **A-D,** The decoding accuracies are baseline corrected by subtracting the 50% chance level. The black dashed vertical lines indicate the onset of stimulus presentation, and the black dashed horizontal lines indicate the chance level of no experimental effect. In gray is the area between the noise ceiling lower and upper bounds.

## Supplementary Figure 3: Encoding models' prediction accuracy (partial correlation)

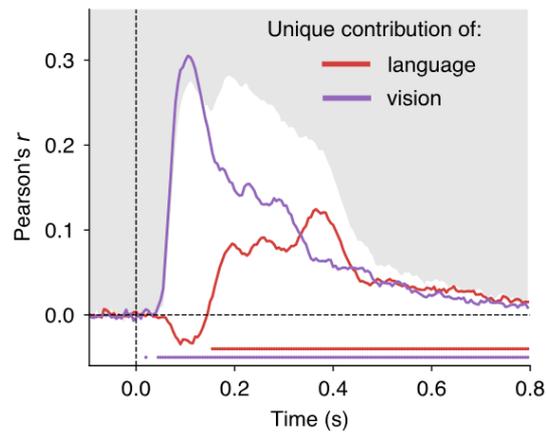

Partial correlation time courses between the recorded EEG test responses and the predicted EEG test responses from the fusion encoding model, controlling for the variance explained by the predicted EEG test responses from either the vision or language encoding models, indicating the unique prediction accuracy contribution of the vision DNN. The black dashed vertical lines indicate the onset of stimulus presentation, and the black dashed horizontal lines indicate the chance level of no experimental effect. In gray is the area between the noise ceiling lower and upper bounds. Colored dots below plots indicate significant points (one sided *t*-test, $p < 0.05$, FDR-corrected for 180 time points, $N = 10$ participants).

# Supplementary Figure 4: Encoding models' prediction accuracy (variance partitioning)

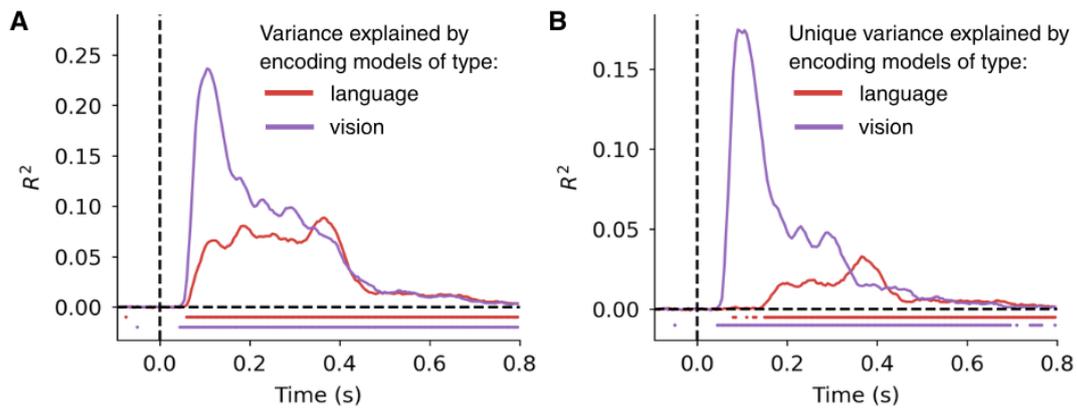

We used variance partitioning to assess the unique variance of EEG responses uniquely predicted by either the vision DNN or the LLM. We trained the vision, language, and fusion encoding models, we computed their explained variance ($R^2$) using the test split, and we adjusted the resulting explained variance scores using the formula $R^2_{adj} = 1 - [\frac{(1-R^2)(n-1)}{(n-k-1)}]$, where $R^2$ is the unadjusted coefficient of determination, $n$ is the number of test split samples, and $k$ is the number of predictors in the model. To compute the unique variance explained by the vision encoding model we subtracted the $R^2_{adj}$ of the language encoding model from the $R^2_{adj}$ of the fusion encoding model. Similarly, to compute the unique variance explained by the language encoding model we subtracted the $R^2_{adj}$ of the vision encoding model from the $R^2_{adj}$ of the fusion encoding model. **A,** Variance explained by the vision, language, and fusion encoding models. **B,** Unique variance explained by the vision and language encoding models. **A-B,** The black dashed vertical lines indicate the onset of stimulus presentation, and the black dashed horizontal lines indicate the chance level of no experimental effect. Colored dots below plots indicate significant points (one sided $t$-test, $p < 0.05$, FDR-corrected for 180 time points, $N = 10$ participants).

# Supplementary Figure 5: Topography plots of encoding models' prediction accuracy (correlation)

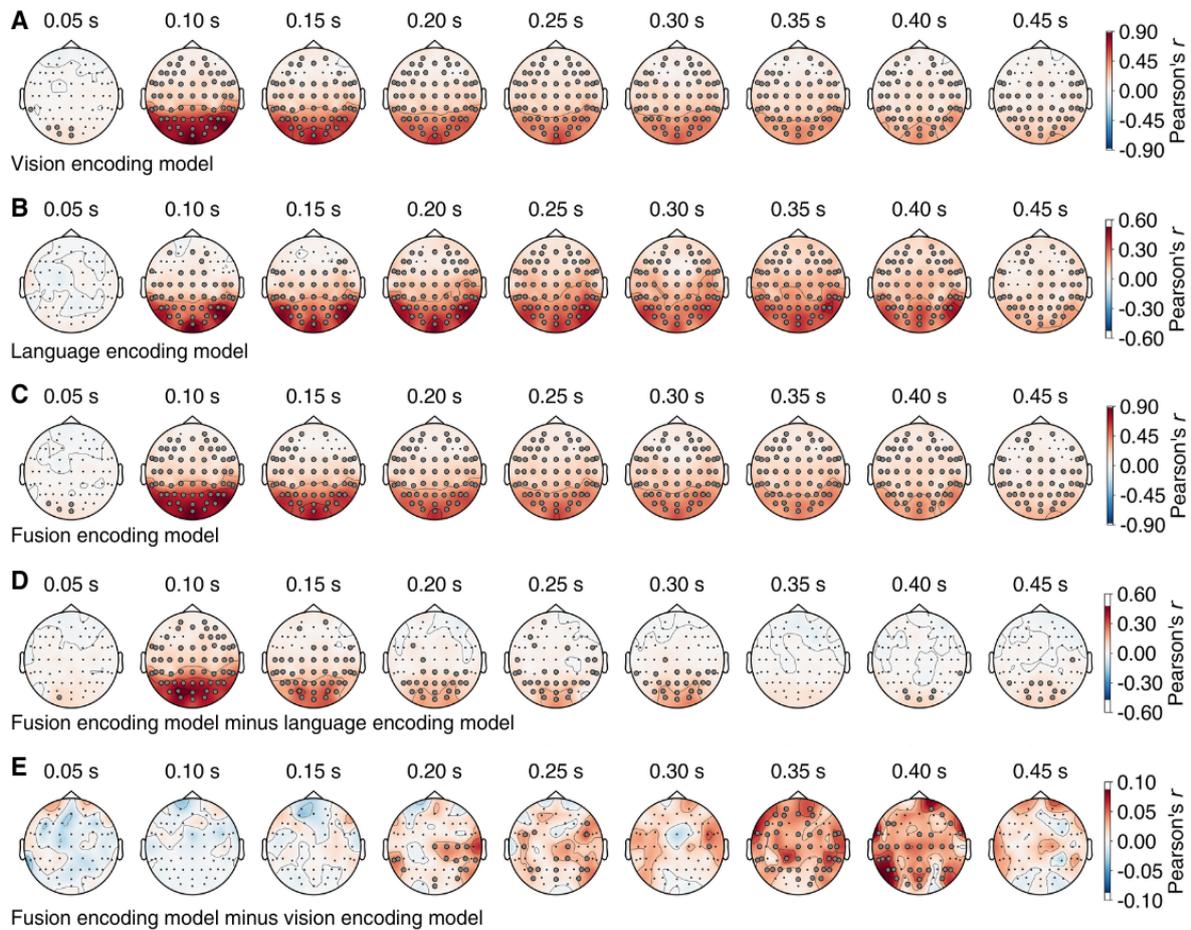

**A-C,** Prediction accuracy (Pearson's *r*) topoplots over time, averaged across participants. **A,** Vision encoding model. **B,** Language encoding model. **C,** Fusion encoding model. **D,** Difference in prediction accuracy between the fusion and the language encoding models, which isolated the effect of the vision DNN. **E,** Difference in prediction accuracy between the fusion and the vision encoding models, which isolated the effect of the LLM. **A-E,** The highlighted black dots indicate significant channels (one-sided *t*-test, $p < 0.05$, FDR-corrected across 180 time points, $N = 10$ participants).

## Supplementary Figure 6: Topography plots of encoding models' prediction accuracy (variance partitioning)

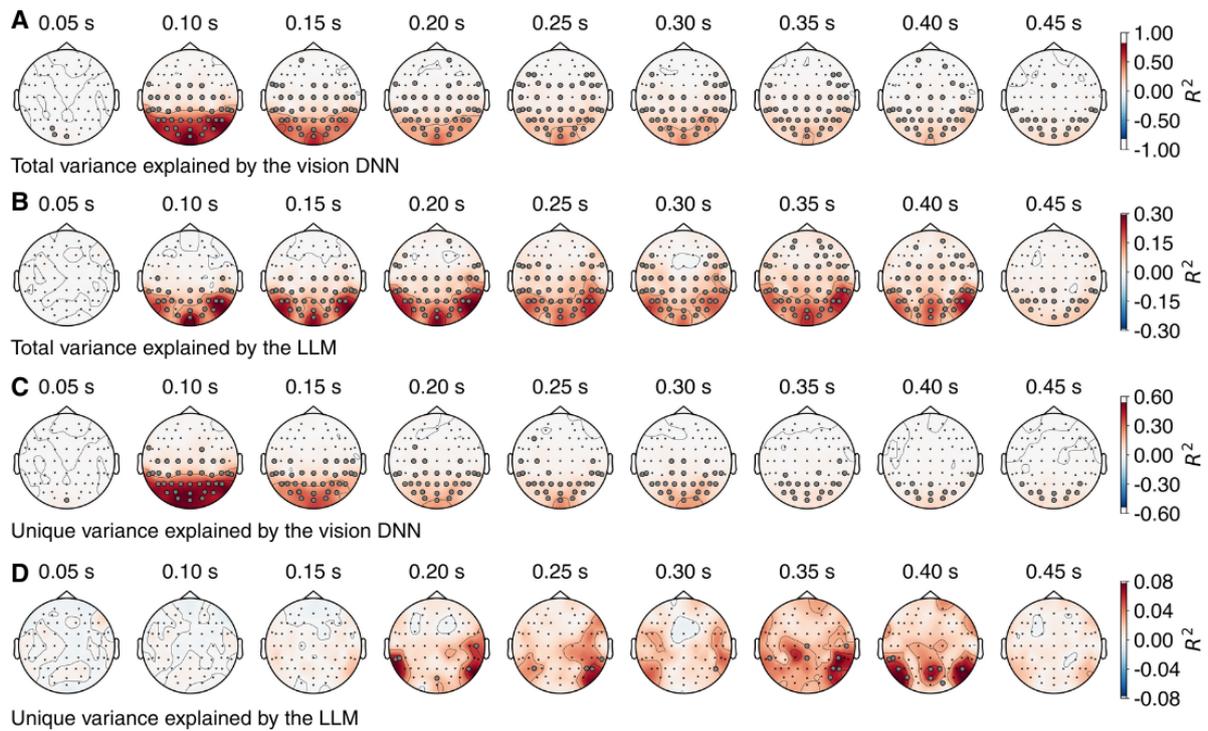

**A**, Variance of the recorded EEG responses explained by vision encoding model. **B**, Variance of the recorded EEG responses explained by the language encoding model. **C**, Unique variance of the recorded EEG responses explained by the fusion encoding model over the language encoding model (thus isolating the effect of the vision DNN). **D**, Unique variance of the recorded EEG responses explained by the fusion encoding model over the vision encoding model (thus isolating the effect of the LLM). **A-D,** Results are averaged across participants. The black dashed vertical lines indicate the onset of stimulus presentation, and the black dashed horizontal lines indicate the chance level of no experimental effect. Colored dots below plots indicate significant points (one sided *t*-test, $p < 0.05$, FDR-corrected for 180 time points, $N = 10$ participants). The highlighted black dots indicate significant channels (one-sided *t*-test, $p < 0.05$, FDR-corrected across 63 channels and 180 time points, $N = 10$ participants).

# Supplementary Figure 7: Prediction accuracy of fusion encoding models trained using different linguistic input

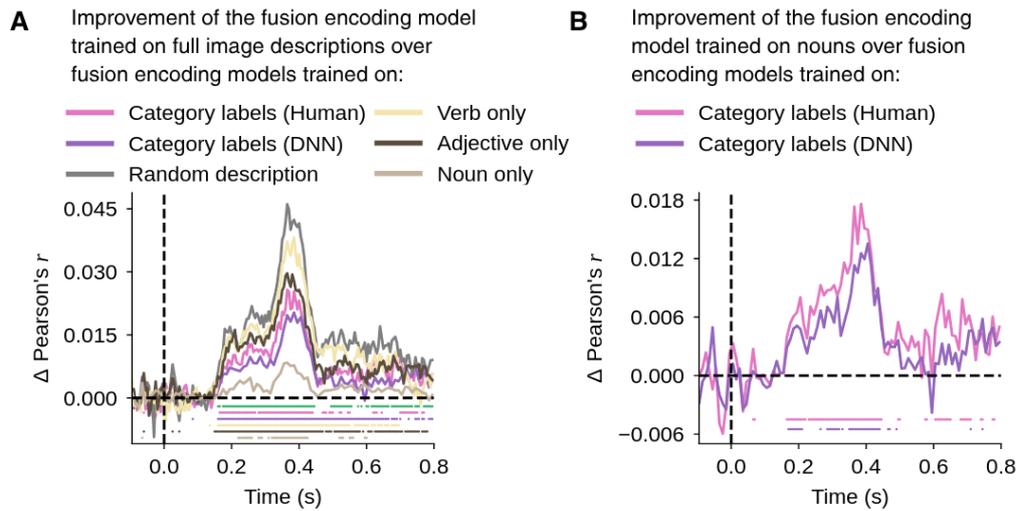

**A,** Difference in prediction accuracy (Pearson's *r*) between the fusion encoding model trained on full image descriptions, and the fusion encoding models trained on object category labels, nouns, adjectives, verbs, and image descriptions randomly assigned to stimulus images. **B,** Difference in prediction accuracy between the fusion encoding model trained on the full image description nouns, and the fusion encoding models trained on object category labels. **A-B,** The black dashed vertical lines indicate the onset of stimulus presentation, and the black dashed horizontal lines indicate the chance level of no experimental effect. Rows of asterisks at the bottom of the plots indicate significant time points (one-sided *t*-test, $p < 0.05$, FDR corrected across 180 time points, $N = 10$ participants).

**Supplementary Figure 8: Time-frequency EEG prediction accuracy (partial correlation)**

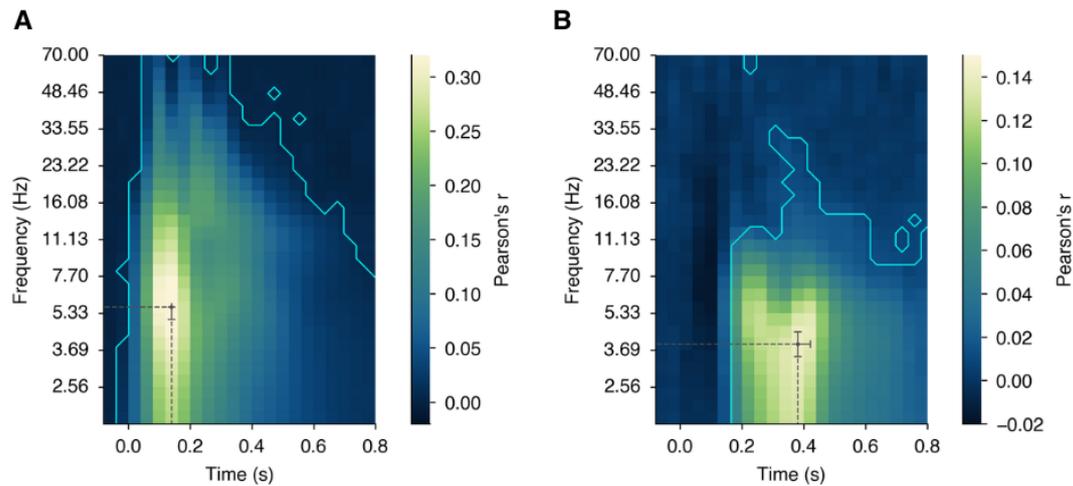

**A**, Partial correlation between the recorded EEG test responses and the predicted EEG test responses from the fusion encoding model, controlling for the variance explained by the predicted EEG test responses from the language encoding model (thus isolating the effect of the vision DNN). **B**, Partial correlation between the recorded EEG test responses and the predicted EEG test responses from the fusion encoding model, controlling for the variance explained by the predicted EEG test responses from the vision encoding model (thus isolating the effect of the LLM). **A-B**, The dashed gray lines indicate latency and frequency peaks of prediction accuracy. Cyan contour lines delineate significant effects (one-sided *t*-test, $p < 0.05$, FDR corrected across all time and frequency points, $N = 10$ participants).

**Supplementary Figure 9: Time-frequency EEG prediction accuracy (variance partitioning)**

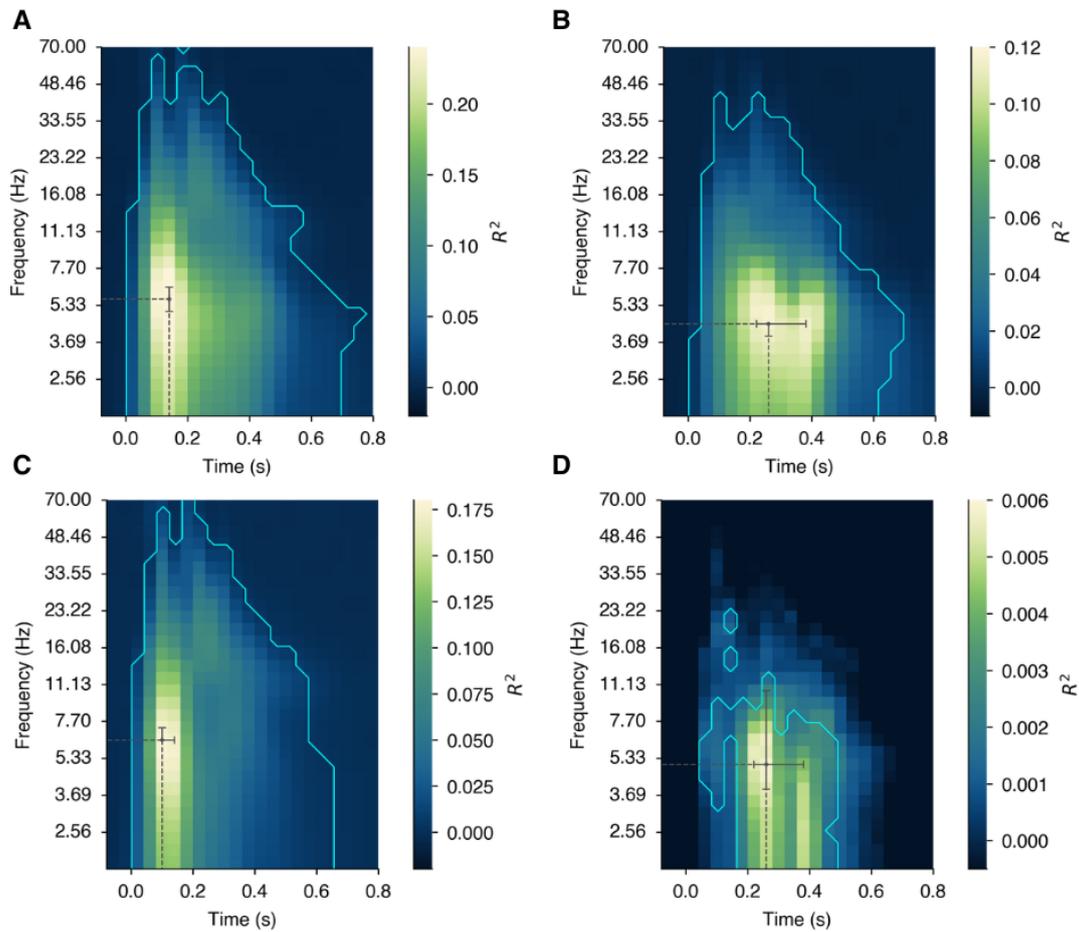

**A**, Variance of the recorded EEG responses explained by vision encoding model **B**, Variance of the recorded EEG responses explained by the language encoding model. **C**, Unique variance of the recorded EEG responses explained by the fusion encoding model over the language encoding model (thus isolating the effect of the vision DNN). **D**, Unique variance of the recorded EEG responses explained by the fusion encoding model over the vision encoding model (thus isolating the effect of the LLM). **A-D**, The dashed gray lines indicate latency and frequency peaks of prediction accuracy. Cyan contour lines delineate significant effects (one-sided *t*-test, $p < 0.05$, FDR corrected across all time and frequency points, $N = 10$ participants).

**Supplementary Figure 10: Time-frequency EEG prediction accuracy (partial correlation; unique contribution of the vision DNN)**

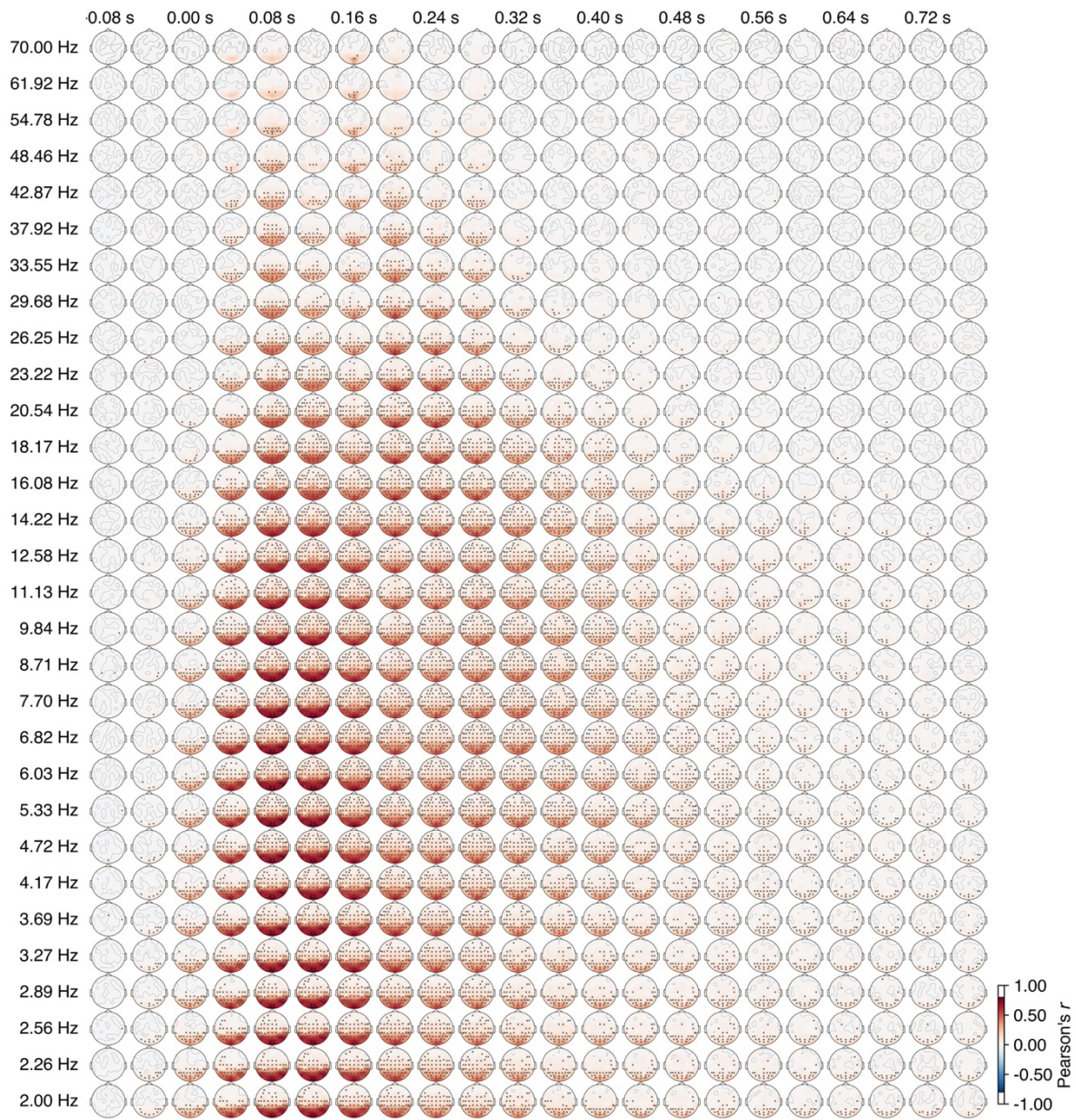

EEG topography of partial correlation results, indicating the unique prediction accuracy contribution of the vision DNN. The highlighted black dots indicate significant channels (one-sided *t*-test, *p* < 0.05, FDR-corrected across 63 channels and 180 time points, *N* = 10 participants).

**Supplementary Figure 11: Time-frequency EEG prediction accuracy (partial correlation; unique contribution of the LLM)**

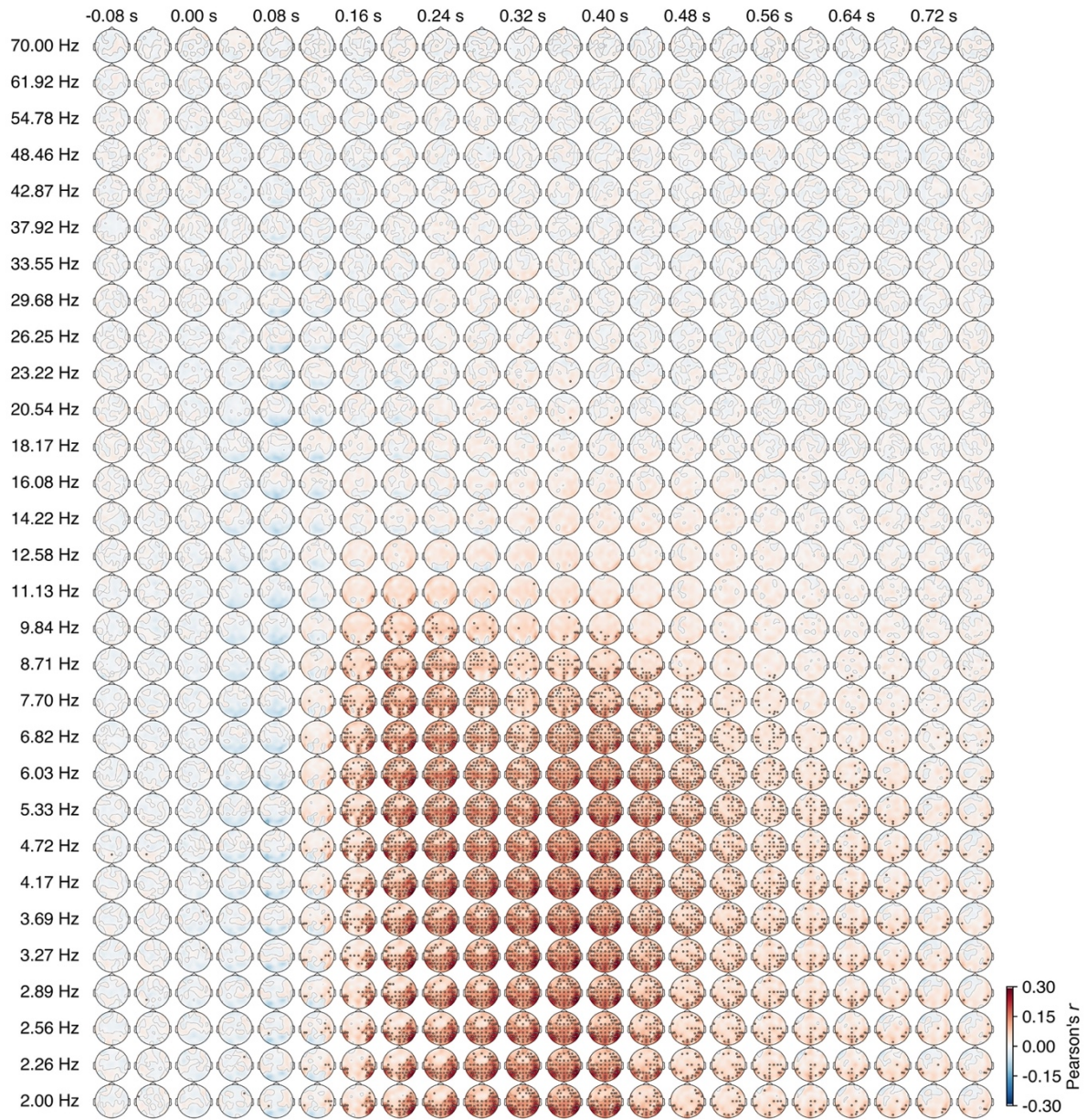

EEG topography of partial correlation results, indicating the unique prediction accuracy contribution of the LLM. The highlighted black dots indicate significant channels (one-sided *t*-test, $p < 0.05$, FDR-corrected across 63 channels and 180 time points, $N = 10$ participants).

**Supplementary Figure 12: Time-frequency EEG prediction accuracy (variance partitioning; unique variance explained by the vision DNN)**

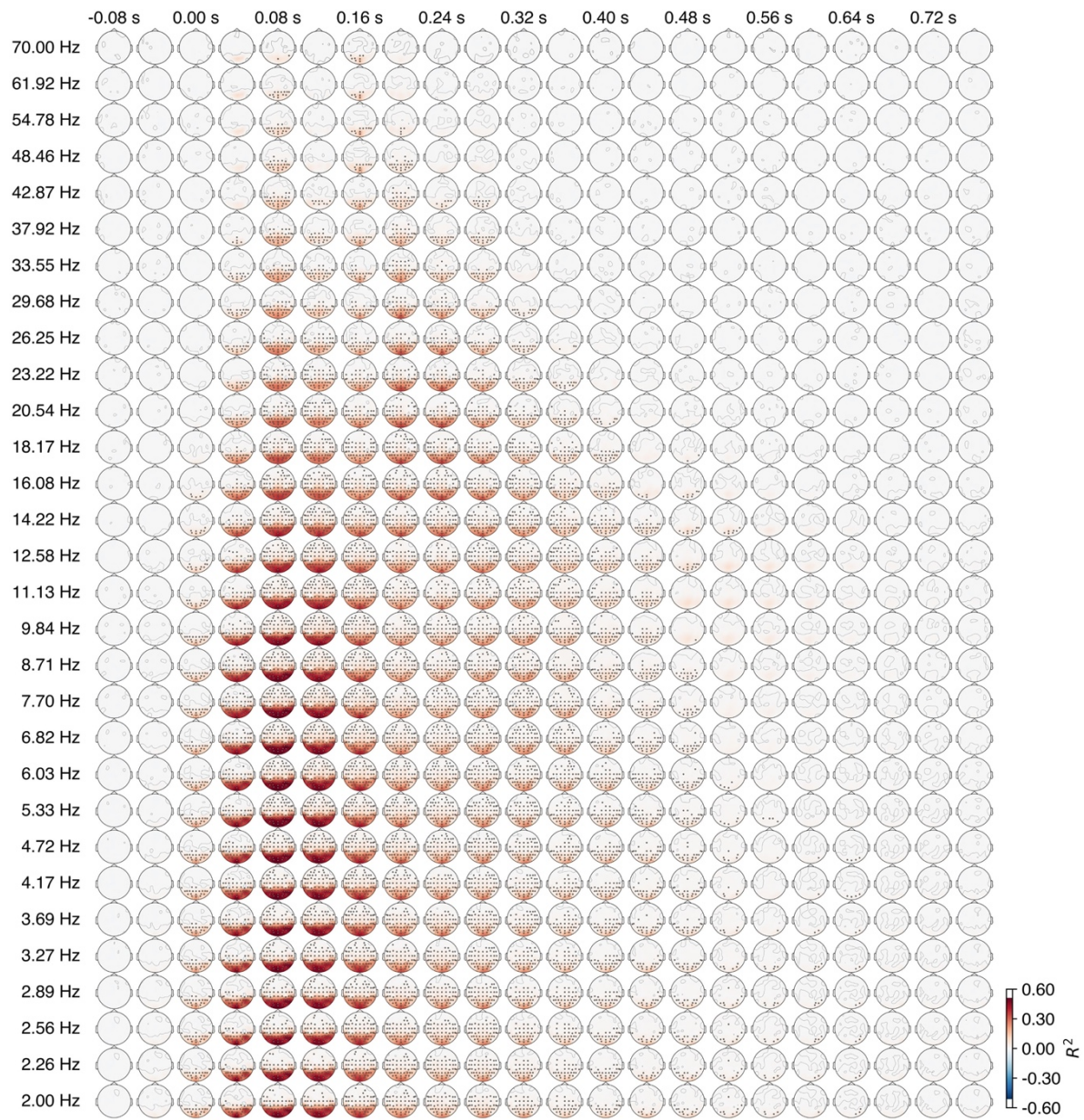

EEG topography of variance partitioning results, indicating the variance uniquely explained by the vision DNN. We performed the variance partitioning analysis using the same rationale detailed in **Suppl. Fig. 4**. The highlighted black dots indicate significant channels (one-sided *t*-test, $p < 0.05$, FDR-corrected across 63 channels and 180 time points, $N = 10$ participants).

# Supplementary Figure 13: Time-frequency EEG prediction accuracy (variance partitioning; unique variance explained by the LLM)

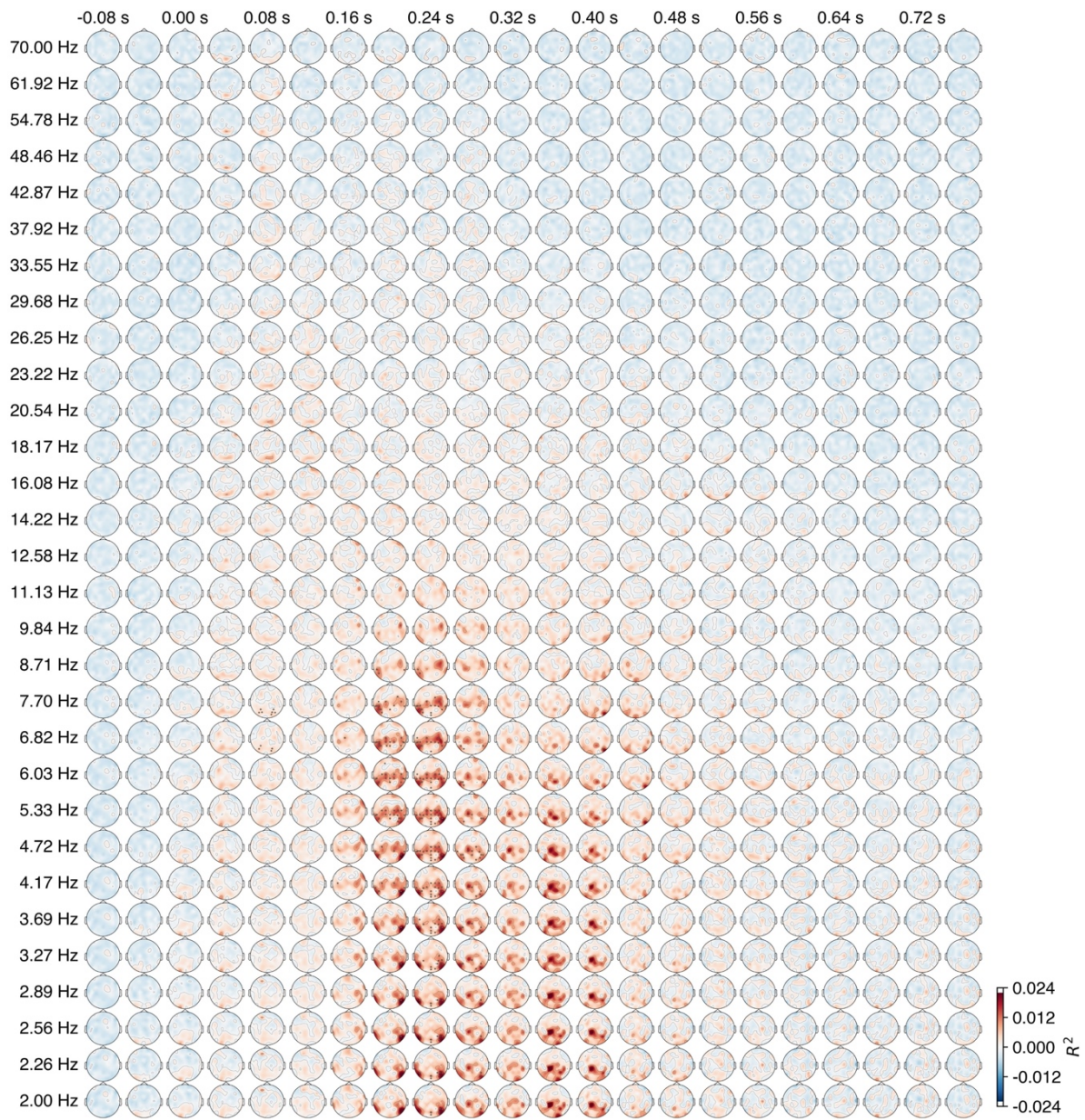

EEG topography of variance partitioning results, indicating the variance uniquely explained by the LLM. We performed the variance partitioning analysis using the same rationale detailed in **Suppl. Fig. 4**. The highlighted black dots indicate significant channels (one-sided $t$-test, $p < 0.05$, FDR-corrected across 180 time points, $N = 10$ participants).

**Supplementary Figure 14: Time-frequency EEG prediction accuracy (subtraction; improvement of the vision DNN over the LLM)**

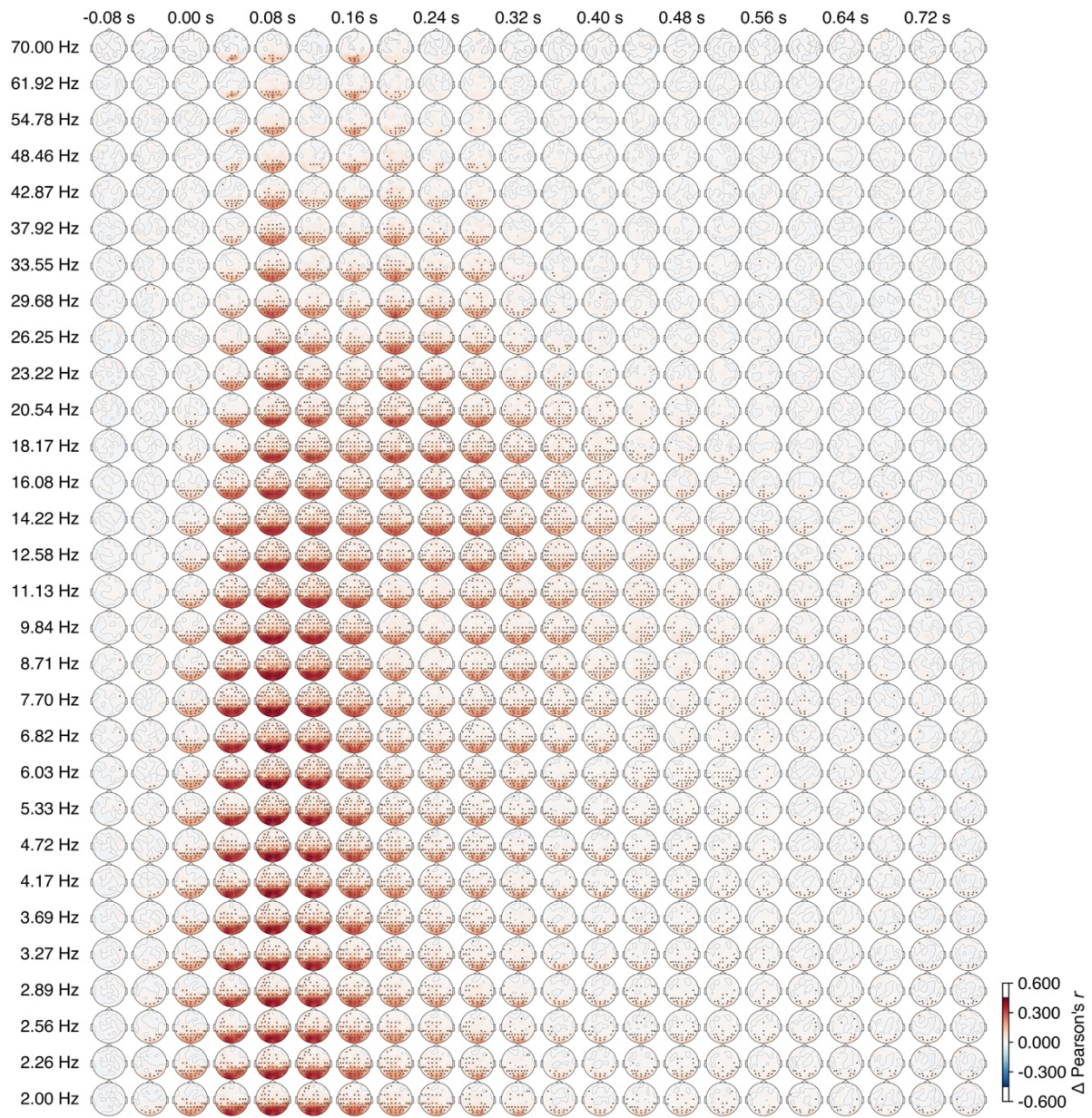

EEG topography of the difference between the fusion and language encoding models prediction accuracies, indicating the improvement of the vision DNN over the LLM. The highlighted black dots indicate significant channels (one-sided *t*-test, $p < 0.05$, FDR-corrected across 63 channels and 180 time points, $N = 10$ participants).

**Supplementary Figure 15: Time-frequency EEG prediction accuracy (subtraction; improvement of the LLM over the vision DNN)**

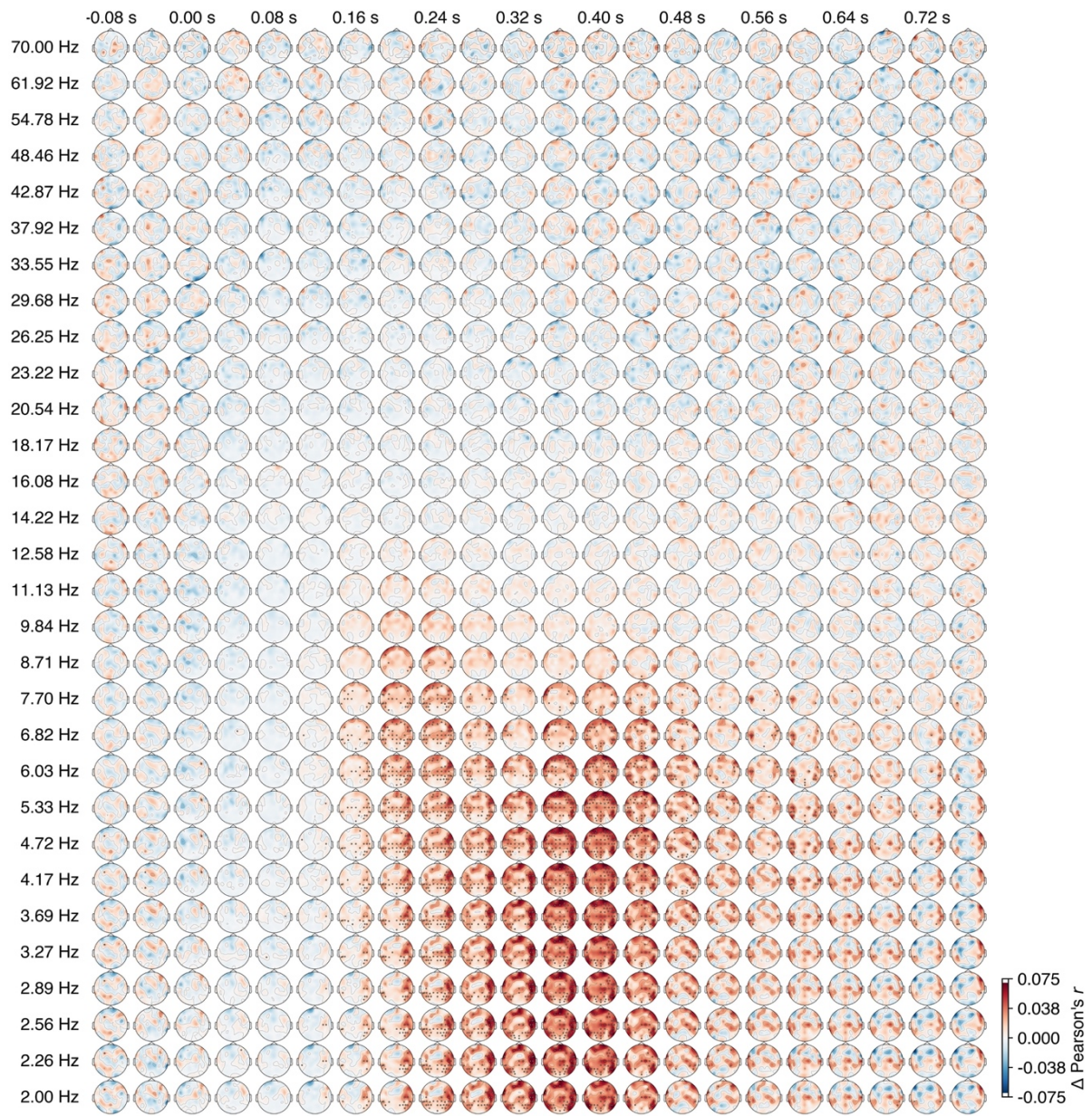

EEG topography of the difference between the fusion and vision encoding models prediction accuracies, indicating the improvement of the LLM over the vision DNN. The highlighted black dots indicate significant channels (one-sided $t$-test, $p < 0.05$, FDR-corrected across 63 channels and 180 time points, $N = 10$ participants).

# Supplementary Tables

## Supplementary Table 1: Bootstrapped onset and peak latency

### A. Bootstrapped onset and peak latency for main conditions

| Method | Encoding model | Onset latency (ms) | Peak latency (ms) |
|---|---|---|---|
| Correlation | Vision | 40 (15-45) | 110 (105-115) |
| | Language | 50 (10-60) | 365 (185-370) |
| | Fusion | 45 (10-45) | 110 (105-115) |
| Subtraction | Fusion minus language | 40 (15-45) | 90 (90-105) |
| | Fusion minus vision | 160 (150-165) | 365 (360-400) |
| Partial correlation | Fusion (vision) | 35 (5-45) | 90 (85-90) |
| | Fusion (language) | 145 (140-155) | 365 (360-400) |
| Variance explained | Vision | 35 (35-45) | 105 (105-115) |
| | Language | 50 (25-60) | 365 (180-370) |
| Unique variance | Vision | 40 (35-45) | 90 (90-110) |
| | Language | 300 (300-300) | 365 (365-370) |

### B. Bootstrapped onset and peak latency difference

| Method | Encoding model 1 | Encoding model 2 | Onset latency difference (ms) | Peak latency difference (ms) |
|---|---|---|---|---|
| Correlation | Vision | Language | 10 (-25–45) | 255 (75–265) *** |
| | Fusion | Language | 5 (-25–45) | 255 (75–265) *** |
| | Vision | Fusion | 5 (-20–25) | 0 (0–5) |
| Subtraction | Fusion minus language | Fusion minus vision | 120 (105–145) **** | 275 (265–310) **** |
| Partial correlation | Fusion (vision) | Fusion (language) | 110 (100–140) *** | 275 (275–310) **** |
| Variance explained | Vision | Language | 15 (-5–25) | 260 (70–265) *** |
| Unique variance | Vision | Language | 260 (255–265) **** | 275 (255–280) **** |

All differences are calculated using model 2 minus model 1

The model names in the partial correlation, e.g. "Fusion (vision)", indicate the unique contribution of the model component within the parentheses.

**** indicate significant at $p < 0.0001$.

*** indicate significant at $p < 0.001$.

** indicate significant at $p < 0.01$.

* indicates significant at $p < 0.05$.

Supplementary Table 2: Bootstrapped onset and peak latency comparison across different encoding model types

| Encoding model | Onset latency (ms) | Peak latency (ms) |
| --- | --- | --- |
| **A. Bootstrapped peak latency for linguistic factor comparison** | | |
| Full description | 45 (10-45) | 110 (105-115) |
| Category label (human) | 45 (20-45) | 110 (105-115) |
| Category label (DNN) | 15 (5-40) | 105 (105-115) |
| Random description | 40 (35-45) | 110 (105-115) |
| Noun only | 40 (-5-45) | 105 (105-115) |
| Adjective only | 45 (25-50) | 110 (105-115) |
| Verb only | 35 (0-45) | 110 (105-115) |
| Full description minus vision | 160 (150-170) | 365 (360-400) |
| Category label (human) minus vision | 180 (155-295) | 380 (355-380) |
| Category label (DNN) minus vision | 160 (155-230) | 365 (345-395) |
| Random description minus vision | - | -30 (-100-45) |
| Noun only minus vision | 160 (-70-170) | 385 (355-395) |
| Adjective only minus vision | 230 (230-345) | 385 (-5-410) |
| Verb only minus vision | - | 0 (-85-750) |
| Full description minus category label (human) | 150 (145-170) | 365 (365-400) |
| Full description minus category label (DNN) | 150 (115-165) | 385 (365-405) |
| Full description minus random description | 145 (130-155) | 365 (365-400) |
| Full description minus verb only | 85 (80-160) | 385 (365-400) |
| Full description minus adjective only | 145 (120-150) | 365 (360-380) |
| Full description minus noun only | 155 (25-270) | 365 (355-385) |
| **B. Bootstrapped onset and peak latency for multimodal model comparison** | | |
| Fusion model | 45 (10-45) | 110 (105-115) |
| CLIP image encoder | 20 (10-45) | 110 (105-115) |
| CLIP text encoder | 55 (45-60) | 365 (115-370) |
| CLIP fusion | 45 (15-50) | 110 (105-120) |
| VisualBERT | 50 (15-60) | 365 (180-370) |
| Fusion minus CLIP image encoder | 55 (40-65) | 365 (350-395) |
| Fusion minus CLIP text encoder | 25 (15-45) | 90 (85-100) |
| Fusion minus CLIP fusion | 25 (5-50) | 100 (90-205) |
| Fusion minus VisualBERT | 45 (5-50) | 90 (85-100) |
| **C. Bootstrapped onset and peak latency for vision DNN comparison** | | |
| CORnet (fusion) | 45 (10-45) | 110 (105-115) |
| AlexNet (fusion) | 45 (0-55) | 110 (105-115) |
| MoCo (fusion) | 45 (15-45) | 110 (105-120) |
| ResNet-50 (fusion) | 45 (15-45) | 110 (105-120) |
| CORnet (fusion minus vision) | 160 (150-170) | 365 (360-400) |
| AlexNet (fusion minus vision) | 150 (145-170) | 370 (350-400) |
| MoCo (fusion minus vision) | 150 (145-160) | 370 (360-400) |
| ResNet-50 (fusion minus vision) | 150 (145-160) | 370 (360-400) |

| D. Bootstrapped onset and peak latency for LLM comparison | | |
|---|---|---|
| text-embedding-3-large (fusion) | 45 (10-45) | 110 (105-115) |
| all-MiniLM-L12-v2 (fusion) | 35 (-25-45) | 110 (100-120) |
| stella-en-1.5B-v5 (fusion) | 45 (-25-45) | 110 (100-120) |
| NV-Embed-v2 (fusion) | 45 (0-45) | 110 (100-120) |
| text-embedding-3-large (fusion minus vision) | 150 (150-165) | 365 (360-400) |
| all-MiniLM-L12-v2 (fusion minus vision) | 150 (140-165) | 370 (360-400) |
| stella-en-1.5B-v5 (fusion minus vision) | 160 (155-235) | 370 (360-400) |
| NV-Embed-v2 (fusion minus vision) | 160 (150-185) | 370 (350-400) |

# Supplementary Table 3: Bootstrapped peak latency and peak frequency of prediction accuracy in the EEG time-frequency domain

**A. Bootstrapped peak latency and frequency in the time-frequency domain**

| Method | Encoding model | Peak latency (ms) | Peak frequency (Hz) |
|---|---|---|---|
| Correlation | Vision | 120 (120-120) | 6.03 (5.33-6.82) |
|  | Language | 240 (200-360) | 4.72 (4.72-5.33) |
|  | Fusion | 120 (120-120) | 6.03 (5.33-6.82) |
| Subtraction | Fusion minus language | 80 (80-80) | 6.82 (6.82-7.70) |
|  | Fusion minus vision | 360 (360-400) | 2.26 (2.00-4.72) |
| Partial correlation | Fusion (vision) | 120 (120-120) | 6.03 (6.03-6.82) |
|  | Fusion (language) | 360 (360-400) | 4.17 (3.69-4.72) |
| Variance explained | Vision | 120 (120-120) | 6.03 (5.33-6.82) |
|  | Language | 240 (200-360) | 4.72 (4.72-5.33) |
| Unique variance | Vision | 80 (80-120) | 6.82 (6.03-6.82) |
|  | Language | 240 (200-360) | 5.33 (2.56-6.82) |

**B. Bootstrapped peak latency and frequency difference**

| Method | Encoding model 1 | Encoding model 2 | Peak latency difference (ms) (model 2 minus model 1) | Peak frequency difference (Hz) (model 1 minus model 2) |
|---|---|---|---|---|
| Correlation | Vision | Language | 120 (80–240) **** | 1.31 (0.62–2.10) ** |
|  | Fusion | Language | 120 (80–240) **** | 1.31 (0.62–2.10) ** |
|  | Vision | Fusion | 0 (0–0) | 0.00 (0.00–0.00) |
| Subtraction | Fusion minus language | Fusion minus vision | 280 (280–320) **** | 4.55 (2.10–5.70) **** |
| Partial correlation | Fusion (vision) | Fusion (language) | 240 (240–280) **** | 1.86 (1.31–3.12) **** |
| Variance explained | Vision | Language | 120 (80–240) **** | 1.31 (0.00–2.10) * |
| Unique variance | Vision | Language | 120 (80–240) **** | 1.31 (0.00–2.10) * |

The model names in the partial correlation, e.g. "Fusion (vision)", indicate the unique contribution of the model component within the parentheses.

**** indicate significant at $p < 0.0001$.

** indicate significant at $p < 0.01$.

* indicates significant at $p < 0.05$.